\documentclass[12pt]{article}
\pdfoutput=1
\usepackage{tikz}
\usetikzlibrary{matrix,arrows,positioning,shapes,snakes,fadings,decorations.pathmorphing,
decorations.pathreplacing,automata,shadows,decorations.markings,patterns}
\usepackage[thicklines]{cancel}
\usetikzlibrary{calc,shapes.callouts,shapes.arrows}
 
\usepackage{mathtools}
\usepackage{jheppub}
\usepackage{amsmath,amssymb,euscript,array,mathrsfs,appendix,ctable,marvosym,bbm,enumitem}
\usepackage{arydshln}
\usepackage{todonotes}
\usepackage{graphicx}
\usepackage{float}
\usepackage[margin=1cm,font=small]{caption}
\usepackage[margin=.2cm,font=footnotesize]{subcaption}
\usepackage[normalem]{ulem}
\usepackage{cleveref}
\usepackage[bottom]{footmisc}

\newcommand\blank[1]{#1}
\renewcommand\blank[1]{}
\def\Buildrel#1\over#2\under#3{\mathrel{\mathop{\kern0pt
#2}\limits^{#1}_{#3}}}

\usepackage{enumitem}

\newcommand{\Tr}{\operatorname{Tr}}

\def\B0{{\boldsymbol 0}}

\def\det{{\rm det}}

\def\ee{\boldsymbol{e}}

\def\Dbarslash{\,\,{\raise.15ex\hbox{/}\mkern-12mu {\bar D}}}
\def\Dslash{\,\,{\raise.15ex\hbox{/}\mkern-12mu D}}
\def\delslash{\,\,{\raise.15ex\hbox{/}\mkern-9mu \partial}}
\def\delbarslash{\,\,{\raise.15ex\hbox{/}\mkern-9mu {\bar\partial}}}

\def\be{\begin{equation}}
\def\ee{\end{equation}}

\DeclareFontFamily{U}{mathx}{\hyphenchar\font45}
\DeclareFontShape{U}{mathx}{m}{n}{
      <5> <6> <7> <8> <9> <10>
      <10.95> <12> <14.4> <17.28> <20.74> <24.88>
      mathx10
      }{}
\DeclareSymbolFont{mathx}{U}{mathx}{m}{n}
\DeclareFontSubstitution{U}{mathx}{m}{n}
\DeclareMathAccent{\widecheck}{0}{mathx}{"71}
\DeclareMathAccent{\wideparen}{0}{mathx}{"75}

\title{Integrable asymmetric $\lambda$-deformations}
 \author[a,b]{Sibylle Driezen,}
   \author[b]{Alexander Sevrin,}
\author[a,b]{ and Daniel C. Thompson}
\affiliation[a]{Department of Physics, Swansea University\\Singleton Park, Swansea SA2 8PP, U.K.}
\affiliation[b]{Theoretische Natuurkunde, Vrije Universiteit Brussel\\  \& The International Solvay Institutes\\
Pleinlaan 2, B-1050 Brussels, Belgium } 

\emailAdd{Sibylle.Driezen@vub.be}{
 \emailAdd{Alexandre.Sevrin@vub.be}
 \emailAdd{D.C.Thompson@swansea.ac.uk}

\abstract{ We  construct integrable deformations of the $\lambda$-type for asymmetrically gauged WZW models.   This is achieved by a modification of the Sfetsos gauging procedure to account for a possible automorphism that is allowed in $G/G$ models.  We verify classical integrability, derive the one-loop beta function for the deformation parameter and give the construction of integrable D-brane configurations in these models.  As an application, we detail the case of the $\lambda$-deformation of the cigar geometry corresponding to the axial gauged $SL(2,R)/U(1)$ theory at large $k$. Here we also exhibit a range of both A-type and B-type integrability preserving D-brane configurations. 
}

\setlength{\parskip}{10pt}

\begin{document}
%

\maketitle

\section{Introduction}
 
Since the observation of worldsheet integrability in the $AdS_5 \times S^5$  superstring  \cite{Bena:2003wd}, integrable two-dimensional non-linear sigma-models have played a prominent role    in the gauge-gravity correspondence. In the planar limit in particular, the simplicity offered by  integrability allows one to go beyond perturbation theory and interpolate  at finite 't Hooft coupling between known results at both sides of the correspondence (for a review see \cite{Beisert:2010jr,Arutyunov:2009ga}).\\
\indent For the purpose of the present paper, we are interested in the application of bosonic integrable sigma models as building blocks of worldsheet theories\footnote{When supplemented with a fermionic field content, as in a Green-Schwarz formulation for instance, they should describe consistent string configurations.} describing strings  propagating in  curved backgrounds. Well known examples in this context are the Wess-Zumino-Witten  (WZW) model \cite{Witten:1983ar}, which has an exact worldsheet CFT formulation, and the Principal Chiral Model (PCM) \cite{Polyakov:1975rr}, which has worldsheet integrability, on a non-Abelian group manifold. Closely related are the gauged WZW model and the Symmetric Space Sigma Model (SSSM) which can be obtained by gauging  an appropriate subgroup of the global symmetry group.  These gauged theories retain some desirable properties;  the gauged WZW model gives a Lagrangian description of coset CFT's \cite{Gawedzki:1988hq,Karabali:1988au} and the SSSM  retains integrability \cite{Eichenherr:1979ci}. Both provide  highly symmetrical target spaces which have been key in the construction of amenable string duals.

\indent An interesting question in recent years has been to deform  known holographic theories while maintaining worldsheet integrability\footnote{One ambition here is to have gravity duals that reduce the amount of (super)symmetries on the gauge theory side as in e.g. \cite{Lunin:2005jy}.}. Prominent examples include the $\eta$- \cite{Klimcik:2008eq,Delduc:2013fga,Delduc:2013qra}, $\beta$- \cite{Lunin:2005jy,Kawaguchi:2014qwa,Osten:2016dvf}\footnote{See also the recent \cite{Borsato:2018spz} and references therein.}  and $\lambda$-deformations \cite{Sfetsos:2013wia,Hollowood:2014rla,Hollowood:2014qma}. Our focus will be on the $\lambda$-deformation  which is an integrable two-dimensional QFT for all values $ \lambda \in [0,1]$. For $\lambda \rightarrow 0$ the model traces back to the WZW model (or gauged WZW model) while for $\lambda \rightarrow 1$  one finds the non-Abelian T-dual of the PCM (or SSSM). There has been significant evidence from both a worldsheet \cite{Hollowood:2014qma,Appadu:2015nfa} and target space \cite{Borsato:2016zcf,Chervonyi:2016ajp,Borsato:2016ose} perspective that, when applied to super-coset geometries, the $\lambda$-model is a marginal deformation introducing no Weyl anomaly. In \cite{Sfetsos:2014cea,Demulder:2015lva} it was also shown one can promote  bosonic coset $\lambda$-models  to type IIB supergravity backgrounds when a suitable ansatz is made for the RR fields. \\
\indent We will focus our attention here on bosonic coset $\lambda$-deformations of $G/H$ gauged WZW models. A limitation to the standard construction so far is that it is deforming    WZW models where only the vector subgroup is gauged \cite{Sfetsos:2013wia,Hollowood:2014rla}. When the subgroup $H$ is Abelian, however, gauging an axial action in the WZW leads to a  topologically distinct target space \cite{Witten:1991mm,Ginsparg:1992af}. For $H$ non-Abelian,  particular asymmetrical gaugings  can be of interest in the case of  higher rank groups \cite{Witten:1991mm,Bars:1991pt}. The present note will  fill this gap by  deforming  spacetimes obtained from asymmetrically gauged WZW models on a general footing\footnote{Similar ideas of an asymmetric deformation have been developed in \cite{Georgiou:2016zyo,Georgiou:2016urf} where a tensor product of coset manifolds is considered with either different levels or an asymmetrical gauging between the tensor product terms (see also the recently appeared \cite{Georgiou:2018gpe}). The novelty of our approach includes deforming an asymmetric gauging of one factor in the tensor product.}.\\
\indent A physical motivation of this line of study is the two-dimensional Euclidean black hole in string theory \cite{Witten:1991yr,Elitzur:1991cb,Mandal:1991tz} corresponding to the $SL(2,R)/U(1)_k$ WZW model \cite{Witten:1991yr,Rocek:1991vk}. When the gauged $U(1)$ is compact and vector one obtains the so-called trumpet geometry, while for an axial gauging one finds the so-called cigar\footnote{These backgrounds are only valid for large $k$, receiving (quantum) corrections for finite $k$ \cite{Tseytlin:1992ri,Bars:1993zf}.}. Analytical continuation of the Euclidean time gives the Minkowskian black hole   where the trumpet  corresponds to the region within the singularity and the cigar to the region outside the horizon \cite{Witten:1991yr,Dijkgraaf:1991ba}. In particular the cigar approaches asymptotically a flat space cylinder while the tip  describes the  horizon itself. These regions are known to be T-dual \cite{Dijkgraaf:1991ba,Giveon:1991sy,Kiritsis:1991zt,Rocek:1991ps} to the $\mathbb{Z}_k$ orbifold of one another and  are indeed described by an equivalent coset CFT \cite{Dijkgraaf:1991ba}. \\
\indent The stringy origin of a black hole horizon has been an attractive asset for the study of the axial $SL(2,R)/U(1)_k$ WZW. In two target space dimensions the only low energy closed string modes are tachyons  winding around the periodic direction of the cigar. However, when these states enter the region of the horizon  at the tip, winding number conservation breaks,
 leading to the existence of a tachyonic condensate in that region. This has been understood in \cite{Kazakov:2000pm} using the (bosonic) FZZ duality \cite{Fateev:SL,Kazakov:2000pm,Hikida:2008pe}  between the cigar geometry and Sine-Liouville theory where the latter is an interacting  theory in a flat space cylinder geometry. Here it is an exponentially growing potential  that breaks winding conservation explicitly and only allows high momentum tachyon modes to penetrate through the  dual  of the region behind the horizon \cite{Giveon:2015cma}.
The machinery developed in this note allows one to study the effects of the $\lambda$-deformation to the cigar geometry and the Sine-Liouville potential explicitly. At this point the interested reader might be enticed by the success of integrability in going beyond perturbation theory to study quantum gravity  effects associated to the horizon. Moreover, using the large $N$ matrix model description of the cigar through Sine-Liouville theory \cite{Kazakov:2000pm}, this particular application opens the route to a tractable interpretation of the integrable $\lambda$-deformations in holography. 

 In section \ref{s:construction} we develop the $\lambda$-deformation of the asymmetrically gauged WZW model. We show that the model is  classically integrable and that, when the asymmetrical gauging respects the symmetric space decomposition\footnote{It seems only a technical issue to relax this requirement.},  the one-loop beta function of the $\lambda$-parameter match those obtained in the case of symmetric gaugings.  We conclude this section by describing  integrable boundary conditions of the worldsheet theory
where we develop  the method of \cite{Driezen:2018glg} to accommodate for coset spaces and asymmetric gaugings.\\
 \indent We then briefly introduce  the $SL(2,R)/U(1)_k$ WZW and  apply the $\lambda$-deformation to the cigar geometry\footnote{Although the region of the deformed cigar geometry was captured globally in \cite{Sfetsos:2014cea} and can be obtained from analytical continuations of the $SU(2)/U(1)$ case of \cite{Sfetsos:2013wia}, the methodology developed here is more fundamental and, moreover, applicable to a wide range of models.} in section \ref{s:SL2U1}.  To  first order we will see the deformation to explicitly break the axial-vector duality  of the undeformed case. The analysis of our method for the integrable boundary conditions, however, shows the D-brane configurations of \cite{Fotopoulos:2003vc,Ribault:2003ss,Fotopoulos:2004ut,Israel:2004jt,Ribault:2005pq} to persist the deformation albeit with isometries being lost. We  find D1-branes extending to asymptotic infinity, but allowed only at particular angles in the deformed cigar, D0-branes at the tip and D2-branes covering the whole or part of the space. In the undeformed case these branes are distinguished, in the nomenclature of \cite{Maldacena:2001ky}, as the former being of A-type, while the latter two  being of B-type. Finally, after a small review on FZZ duality, we give the starting point to the study of a deformed Sine-Liouville theory by extracting the first order perturbation.\\
 \indent We conclude in section \ref{s:conclusions} with a short summary and outlook of our results.

\section{Left-right asymmetrical $\lambda$-deformations} \label{s:construction}

In this section we  generalise the construction  of $\lambda$-deformations of symmetric coset manifolds $G/H$ developed in \cite{Sfetsos:2013wia,Hollowood:2014rla,Hollowood:2014qma} to incorporate  the possibility of deforming the left-right asymmetrical gauged WZW model \cite{Bars:1991pt,Witten:1991mm}.

This \textit{asymmetric} coset $\lambda$-deformation  is constructed in a number of steps based on the Sfetsos gauging procedure \cite{Sfetsos:2013wia}. First one combines\footnote{For a summary of our conventions and more details on the WZW and SSSM   we refer the reader to the appendix \ref{a:appendix}.  } the Wess-Zumino-Witten (WZW) model \cite{Witten:1983ar} on a group manifold $G$,
\begin{equation}\label{eq:ActionWZW}
S_{\mbox{\tiny{WZW}},k}(g) = - \frac{k}{2\pi} \int_{\Sigma} d\sigma d\tau \langle  g^{-1}\partial_+ g , g^{-1} \partial_- g \rangle - \frac{k}{24\pi} \int_{M_3}  \langle \bar{g}^{-1} \mathrm{d}\bar{g} , \left[  \bar{g}^{-1} \mathrm{d}\bar{g} ,  \bar{g}^{-1} \mathrm{d}\bar{g} \right]  \rangle ,
\end{equation}
 with the Symmetric Space Sigma Model (SSSM) on $G/H$,
 \begin{equation}\label{eq:SSSMaction}
S_{\mbox{\tiny{SSSM}},\kappa^2}(\widehat{g}, B_\pm)  = - \frac{\kappa^2 }{\pi} \int d\sigma d\tau \langle (\widehat{g}^{-1}\partial_+ \widehat{g} - B_+),( \widehat{g}^{-1}\partial_- \widehat{g} - B_-) \rangle ,
\end{equation}
 where the latter is invariant under an $H_R \subset G$ action $\widehat{g} \rightarrow \widehat{g} h$ with $h\in H$ when the gauge fields $B_\pm \in \mathfrak{h}$ transform as $ B_\pm \rightarrow h^{-1} \left( B_\pm  + \partial_\pm \right) h$. Note that these models are  realised through distinct group elements $g \in G$ and $\widehat{g}\in G$ respectively which we assume to be connected to the identity. Next, we reduce back to $\dim G - \dim H$ degrees of freedom by gauging simultaneously the left-right asymmetric $G$-action in the WZW model (generalising the usual $\lambda$-model construction \cite{Sfetsos:2013wia,Hollowood:2014rla,Hollowood:2014qma} where the vector action is gauged) and the $G_L$-action in the SSSM given by,
\begin{equation}
 \begin{aligned}
 g &\rightarrow g_0^{-1} g \widetilde{g}_{0},\label{eq:gaugetransformation}\\
\widehat{g} &\rightarrow g_0^{-1} \widehat{g}.
 \end{aligned}
\end{equation}
 Here $g_0 = \exp(G^A T_A) \in G $ and $ \widetilde{g}_0 = \exp(G^A \widetilde{T}_A) \in G$ have the same parameters $G^A$ but are generated  by different embeddings $T_A$ and $ \widetilde{T}_A$  of a representation of the Lie algebra $\mathfrak{g}$ of $G$. Their relation can be packaged into an object $W$ as $\widetilde{T}_A = W(T_A) = W^{B}{}_A T_B$. To find a gauge-invariant action we introduce the gauge fields $A_\pm = A_\pm^A T_A$ transforming as, 
\begin{equation}
\begin{aligned}
A_\pm \rightarrow g_0^{-1} \left( A_\pm - \partial_\pm \right) g_0 ,  \qquad
 W(A_\pm) \rightarrow \widetilde{g}_0^{-1} ( W(A_\pm) - \partial_\pm ) \widetilde{g}_0  ,
\end{aligned}
\end{equation} 
 and we perform the usual minimal substitution (i.e. replacing derivatives by $ \partial_\pm \cdot - A_\pm \cdot $) in the SSSM term and replace the WZW term by the left-right asymmetrical gauged WZW model\footnote{In the following, we will abbreviate the left-right asymmetrical gauged WZW  model with $G/H_{AS}$ WZW when the subgroup $H \subset G$ is gauged.} \cite{Bars:1991pt,Witten:1991mm} on the coset   $G/G_{AS}$ given by,
\begin{equation}\label{eq:ASGWZW}
\begin{aligned}
S_{\mbox{\tiny{WZW}},k}(g,A^A_\pm,W) ={}& S_{\mbox{\tiny{WZW}},k}(g) + \frac{k}{\pi} \int_\Sigma d\sigma d\tau \langle A_- , \partial_+ g g^{-1} \rangle - \langle W(A_+), g^{-1} \partial_- g  \rangle \\
 &+ \langle A_-, g W(A_+) g^{-1} \rangle - \frac{1}{2}\langle  A_-, A_+ \rangle - \frac{1}{2} \langle W(A_-), W(A_+)  \rangle .
\end{aligned}
\end{equation}
The latter is  gauge-invariant\footnote{The invariance under the gauge transformations \eqref{eq:gaugetransformation} can be easily checked when rewriting the action \eqref{eq:ASGWZW} 	using the Polyakov-Wiegmann identity \cite{Polyakov:1984et}, which in our conventions takes the form,
\begin{equation*}
S_{\mbox{\tiny{WZW}},k}(g_1 g_2 ) = S_{\mbox{\tiny{WZW}},k}(g_1) + S_{\mbox{\tiny{WZW}},k}(g_2 ) - \frac{k}{\pi} \int \mathrm{d}\sigma \mathrm{d}\tau \langle g_1^{-1} \partial_- g_1 , \partial_+ g_2 g_2^{-1} \rangle ,
\end{equation*}
for $g_1 , g_2 \in G$. One obtains $S_{\mbox{\tiny{WZW}},k}(g,A^A_\pm,W) = S_{\mbox{\tiny{WZW}},k}(g_L^{-1}g\tilde g_R)- S_{\mbox{\tiny{WZW}},k}(g_L^{-1} g_R)$, where $g_{L,R}\in G$ and one identifies $A_+= \partial_+ g_R\,g_R^{-1}$ and $A_-= \partial_- g_L\,g_L^{-1}$. The gauge transformations are given by
$g \rightarrow g_0^{-1} g \widetilde{g}_{0}  $ and $g_{L,R} \rightarrow g_0^{-1} g_{L,R}$.
 } provided that $W: \mathfrak{g} \rightarrow \mathfrak{g}$ is a metric-preserving automorphism of the Lie algebra  $\mathfrak{g}$  \cite{Bars:1991pt,Witten:1991mm} i.e.,
\begin{equation}\label{eq:Wconditions}
 W ( \left[ T_A , T_B \right] ) = \left[ W ( T_A ), W(T_B ) \right] \quad \text{and} \quad \langle W(T_A), W (T_B) \rangle = \langle T_A, T_B \rangle .
\end{equation}
Finally, one can fix the gauge symmetry by setting $\widehat{g} = \mathbf{1}$, which allows one to integrate out the gauge fields $B_\pm$ easily. The result is  a generalised version\footnote{When the automorphism $W = \mathbf{1}$ one finds  the usual $\lambda$-model on the $G/H$  coset \cite{Sfetsos:2013wia,Hollowood:2014rla} which is deforming the \textit{vectorially} gauged $G/H_V$ WZW model.} of the $\lambda$-deformed gauged WZW given by,
\begin{equation}\label{eq:ActionLambdaCosetGaugeFields}
\begin{aligned}
S_\lambda (g, A^A_\pm,W ) ={}&S_{\mbox{\tiny{WZW}},k}(g) + \frac{k}{\pi} \int d\sigma d\tau \langle A_-,  \partial_+ g g^{-1} \rangle - \langle W(A_+),g^{-1}\partial_- g \rangle \\
& + \langle A_-, g W(A_+) g^{-1} \rangle - \langle A_+, \Omega (A_-) \rangle ,
\end{aligned}
\end{equation}
where we introduced the operator $\Omega (\mathfrak{g})  =\mathfrak{g}^{(0)} \oplus \frac{1}{\lambda} \mathfrak{g}^{(1)}$ with  $\mathfrak{g}^{(0)} \equiv \mathfrak{h}$. The deformation parameter $\lambda$ is defined as $\lambda = \frac{k}{\kappa^2 + k}$. \\
\indent The action \eqref{eq:ActionLambdaCosetGaugeFields} still has a residual $\dim H$  left-right asymmetrical gauge symmetry  inherited from the $G/G_{AS}$ WZW  model \eqref{eq:ASGWZW} which acts  as,
\begin{equation}\label{eq:residualtransf}
\begin{aligned}
g &\rightarrow h^{-1} g \widetilde{h} , \\
A_\pm^{(0)} &\rightarrow h^{-1} \left( A_\pm^{(0)} - \partial_\pm \right) h , \qquad A_\pm^{(1)} \rightarrow h^{-1}  A_\pm^{(1)} h, 
\end{aligned}
\end{equation}
with  $h = \exp (X)$,  $\widetilde{h} = \exp (W(X))$ connected to the identity and where $X\in \mathfrak{g}^{(0)}$. Consequently under the gauge transformation we have $W(A_\pm^{(0)}) \rightarrow \widetilde{h}^{-1} ( W(A_\pm^{(0)})  - \partial_\pm ) \widetilde{h}$ and  $W(A_\pm^{(1)})  \rightarrow  \widetilde{h}^{-1}  W(A_\pm^{(1)})  \widetilde{h}$. This shows that the fields $A^{(0)}_\pm$ are still genuine (but non-propagating) gauge fields while  the fields $A^{(1)}_\pm$ are auxiliary. Both can be integrated out, yielding the constraints,
\begin{equation} \label{eq:constraints}
\begin{aligned}
A_+ &= - \left( D_g  W - \Omega \right)^{-1} \partial_+ g g^{-1} ,\\
A_- &=  \left(D_{g^{-1}} -  W \Omega \right)^{-1} g^{-1} \partial_- g .
\end{aligned}
\end{equation}
Once  the gauge fields are eliminated in favour of these equations, the resulting action is given by,
\begin{equation}\label{eq:LambdaActionEffective}
\begin{aligned}
S_\lambda (g , W) = S_{\mbox{\tiny{WZW}},k}(g) + \frac{k}{\pi} \int d\sigma d\tau \langle  \partial_+ g g^{-1} , \left( \mathbf{1} - D_g W \Omega \right)^{-1}     \partial_- g g^{-1} \rangle ,
\end{aligned}
\end{equation}
accompanied with a non-constant dilaton profile, coming from the Gaussian integral  over gauge fields, given by,
\begin{equation}\label{eq:LambdaDilaton}
e^{-2 \Phi} = e^{-2 \Phi_0} \, \det \left( D_g W - \Omega \right) ,
\end{equation}
with $\Phi_0$ constant.

In the $\lambda \rightarrow 0 $ limit one  reproduces the $G/H_{AS}$ WZW (i.e. the action \eqref{eq:ASGWZW} but with $A_\pm^{(1)} = 0$) which can be seen directly from the constraint equations. For small $\lambda$ one finds, by integrating out the auxiliary fields $A_\pm^{(1)}$ in \eqref{eq:ActionLambdaCosetGaugeFields},  the first order correction to  the $G/H_{AS}$ WZW to be,
\begin{equation}
S_\lambda (g, A_\pm^{(0)},  W ) = S_{\mbox{\tiny{WZW}},k}(g, A_\pm^{(0)}, W ) + \frac{\lambda}{\pi k} \int d\sigma d\tau \,  \langle  \mathcal{J}^{(1)}_+ , W^{-1}  \mathcal{J}_- \rangle + \mathcal{O}(\lambda^2 ),
\end{equation}
where we introduced the  Kac-Moody currents $\mathcal{J}_\pm$  of the $G/H_{AS}$ WZW\footnote{Although we are not aware of an occurrence in the literature of these currents in the case of the $G/H_{AS}$ WZW, they can be  derived analoguously to \cite{Bowcock:1988xr} showing that their Poisson brackets satisfy two commuting classical versions of a Kac-Moody algebra.} defined as
\begin{equation}\label{eq:gWZWcurrents}
\begin{aligned}
\mathcal{J}_+ = - k ( \partial_+g g^{-1} + g W(A^{(0)}_+) g^{-1} - A^{(0)}_- ), \quad
 \mathcal{J}_- = k ( g^{-1}\partial_- g  - g^{-1} A^{(0)}_- g + W(A^{(0)}_+) ) ,
\end{aligned}
\end{equation}
Hence,  the perturbation term  away from the CFT point  is   a particular coupling between these currents. Under the residual gauge transformation \eqref{eq:residualtransf} the currents transform as,
\begin{equation}
\begin{aligned}
\mathcal{J}_+ \rightarrow h^{-1} \mathcal{J}_+ h + k h^{-1} \partial_\sigma h , \qquad \mathcal{J}_- \rightarrow \widetilde{h}^{-1} \mathcal{J}_- \widetilde{h} - k W( h^{-1} \partial_\sigma h ),
\end{aligned}
\end{equation}
so  that the perturbation term  is  gauge invariant as is indeed required for consistency.\\
\indent Another interesting limit to consider is   the $\lambda \rightarrow 1$ scaling limit (sending  $k \rightarrow \infty$) for which in the usual vectorial gauged case of \cite{Sfetsos:2013wia} one  reproduces the non-Abelian T-dual of the SSSM. This fact can be traced back to the property that the $G/G_V$ WZW under the scaling limit reduces to a Langrange multiplier term. For the $G/G_{AS}$ WZW \eqref{eq:ASGWZW} this is not true for general $W$ which  strongly suggests there is no interpretation of this limit as a non-Abelian T-dual.

The novelty of the constructed coset $\lambda$-model  \eqref{eq:ActionLambdaCosetGaugeFields} is that it deforms the left-right asymmetrically gauged $G/H_{AS}$ WZW model \eqref{eq:ASGWZW}  instead of solely the vectorial gauged $G/H_{V}$ WZW. As advertised, this will allow us to deform also  target spaces obtained by an axial gauging when the subgroup $H$ is abelian.  However, even in the undeformed case, as noted in \citep{Bars:1991pt}, not all $W$ that satisfy the conditions \eqref{eq:Wconditions} will produce interesting and novel spacetimes. Indeed, if $W$ is an inner automorphism of the Lie algebra, where one can always find a constant $w \in G$ so that $W(T_A) = w T_A w^{-1}$, the action \eqref{eq:ActionLambdaCosetGaugeFields} can be rewritten as,
\begin{equation} \label{eq:AsymmetricGaugingFields}
 S_\lambda ( g , A_\pm^A, W ) = S_\lambda ( g w, A_\pm^A, \mathbf{1} ),
 \end{equation} 
where we used the $G_L \times G_R$ invariance of the WZW term. Hence, in this case only a trivial redefinition of the fields $g \in G$ to $g w \in G$ has been performed. Nevertheless, if $w \in G^{\mathbb{C}}$ or a different outer automorphism of the Lie algebra the generalisation is non-trivial as we will see later in section \ref{s:SL2U1}.

To conclude this section, we note that the construction as described above is also applicable to the group manifold and super-coset case. For the former one can perform the gauging procedure starting with a combination of a WZW and an ordinary PCM model on a Lie group $G$. The formulae in this section then continue to persist upon the redefinition  $\Omega  = \lambda^{-1}$. We believe this asymmetrical $\lambda$-model can have an interest for higher rank group manifolds allowing Dynkin outer automorphisms such as for instance when $G=SU(N)$, $N>2$.  Moreover, one can view this $\lambda$-model  as one with a single but anisotropic coupling matrix $\lambda^{AB} = \lambda W^{AB}$ as discussed for instance in \cite{Sfetsos:2015nya,Georgiou:2016urf}.  In the super-coset case, where $G$ is a Lie supergroup, the Sfetsos gauging procedure is not applicable anymore, but one can follow straightforwardly the construction of \cite{Hollowood:2014qma} and replace the $G/G_V$ WZW with the $G/G_{AS}$ WZW. The conditions on the automorphism $W$ are analogous to \eqref{eq:Wconditions} but here the inner product on the Lie supergroup will be taken to be the supertrace $\mathrm{STr}$ instead of an ordinary trace. When, moreover, the Lie superalgebra has a semi-symmetric space decomposition defined by  a $\mathbb{Z}_4$ grading $\mathfrak{g} = \oplus_{i=0}^3 \mathfrak{g}^{(i)}$ where $\mathfrak{g}^{(0)} \equiv \mathfrak{h}$ and $\left[\mathfrak{g}^{(i)}, \mathfrak{g}^{(j)} \right] \subset \mathfrak{g}^{(i+j \text{ mod } 4)}$, the formulae in this section  are again similar upon the redefinition $\Omega (\mathfrak{g}) =  \mathfrak{g}^{(0)} \oplus \lambda^{-1} \mathfrak{g}^{(1)} \oplus \lambda^{-2} \mathfrak{g}^{(2)} \oplus \lambda \mathfrak{g}^{(3)}$ and upon the usage of the supertrace. Note that, with respect to the supertrace, $\Omega$ is not symmetric anymore, so that the constraint equations \eqref{eq:constraints} are however altered as,
\begin{equation}
\begin{aligned}
A_+  &= - \left( D_g W - \Omega^T \right)^{-1} \partial_+ g g^{-1},\\
A_- &= \left( D_{g^{-1}} - W \Omega \right)^{-1} g^{-1} \partial_- g ,
\end{aligned}
\end{equation}
with $\Omega^T(\mathfrak{g}) =  \mathfrak{g}^{(0)} \oplus \lambda \mathfrak{g}^{(1)} \oplus \lambda^{-2} \mathfrak{g}^{(2)} \oplus \lambda^{-1} \mathfrak{g}^{(3)} $.


\subsection{Classical integrability}

To check the integrability of the asymmetrical $\lambda$-model we follow the method of \cite{Hollowood:2014rla}\footnote{Note that to translate  to \cite{Hollowood:2014rla}  one should identify the group fields as $g = \mathcal{F}^{-1}$. The method of \cite{Hollowood:2014rla}  consists of  relating the equations of motions of the fields in the $\lambda$-model to the equations of motions of the SSSM for which the Lax pair is known.} starting from the action \eqref{eq:ActionLambdaCosetGaugeFields}. As in the SSSM it is necessary here to assume the Lie algebra to have a symmetric space decomposition defined by $\mathfrak{g} = \mathfrak{g}^{(0)} \oplus \mathfrak{g}^{(1)}$, with $\mathfrak{g}^{(0)} \equiv \mathfrak{h}$, and a $\mathbb{Z}_2$ grading $[\mathfrak{g}^{(i)}, \mathfrak{g}^{(j)}] \subset \mathfrak{g}^{(i+j \text{ mod }2)}$. \\
\indent The equations of motion of the group fields $g$ can be written as,
 \begin{equation}
 \left[ \partial_+ - W(A_+), \partial_- + g^{-1}\partial_-g - g^{-1}A_- g \right] = 0,
 \end{equation}
 or equivalently,
 \begin{equation}
 \left[ \partial_+ - \partial_+ g g^{-1} - g W(A_+) g^{-1}, \partial_- - A_- \right]  = 0.
 \end{equation}
 Using the constraints \eqref{eq:constraints} and  $W$ being a constant Lie algebra automorphism these can be rewritten as,
 \begin{equation}
 \begin{aligned}
 \left[ \partial_+ - A_+, \partial_- - \Omega(A_-) \right] &=0,\\
 \left[ \partial_+ - \Omega(A_+), \partial_- -A_- \right] &=0.
 \end{aligned}
 \end{equation}
The above equations of motion can be represented through a $\mathfrak{g}^{\mathbb{C}}$-valued Lax connection depending on a spectral parameter $z\in \mathbb{C}$ that  satisfies a zero-curvature condition,
\begin{equation}
 \left[ \partial_+ + \mathcal{L}_+(z) , \partial_- + \mathcal{L}_- (z) \right] = 0, \qquad \forall z \in \mathbb{C},
 \end{equation} 
  when it is given by,
 \begin{equation}\label{eq:cosetlambdalax}
 \mathcal{L}_\pm (z) = - A_\pm^{(0)} - z^{\pm 1} \lambda^{-1/2}A_\pm^{(1)}.
 \end{equation}
 This fact shows  the left-right asymmetrical $\lambda$-theories on $G/H$ manifolds to be classically integrable models  \cite{Zakharov:1973pp} for general automorphisms $W$. These $\lambda$-models therefore supplement the list   of \cite{Georgiou:2016urf} of integrable $\lambda$-models  with a general single coupling matrix  for $\lambda^{\alpha\beta} = \lambda W^{\alpha\beta}$    with $W$ satisfying \eqref{eq:Wconditions}. Additionally, along similar lines, one can show integrability for the asymmetrical $\lambda$-model on group and super-coset manifolds for which the Lax connection will take the form,
 \begin{equation}
  {\cal L}_\pm (z) = - \frac{2}{1+\lambda} \frac{1}{1 \mp z} A_\pm ,
 \end{equation}
 and,
 \begin{equation}
 {\cal L}_\pm (z) = - A_\pm^{(0)} - z^{-1} \lambda^{\pm 1/2} A_\pm^{(1)} - z^{\pm 2} \lambda^{-1} A_\pm^{(2)} - z \lambda^{\mp 1/2} A_\pm^{(3)} ,
 \end{equation}
respectively.

\subsection{One-loop beta functions}

To compute the one-loop beta functions  of the $\lambda$-parameter of the above asymmetrically deformed theories, we follow the method of \cite{Appadu:2015nfa}, but see also \cite{Itsios:2014lca,Sfetsos:2014jfa} for possibly different approaches. The authors of \cite{Appadu:2015nfa} consider fluctuations  around a background field for the currents rather than the fundamental field $g$ and applied the background field approach  to the PCM and the SSSM. They  efficiently generalise their results to the usual $\lambda$-deformed theories on group or (super)-coset manifolds by identifying the appropriate fields such that the classical equations of motion take an identical form to those of the PCM or SSSM models respectively. With minor adjustments we can follow the same path here.

To begin we choose for the  group valued field $g$ the same background  as \cite{Appadu:2015nfa}, namely,
\begin{equation}\label{eq:BGSSSM}
g = \exp \left( \sigma^+ \Lambda_+ + \sigma^- \Lambda_- \right),
\end{equation}
with $\Lambda_\pm$  constant commuting elements of $\mathfrak{g}^{(1)}$. Hence, on the background we have $\partial_\pm g g^{-1} = g^{-1} \partial_\pm g = \Lambda_\pm$. Through the constraints \eqref{eq:constraints} the background  of the gauge fields $A_\pm$ then becomes,
\begin{equation}\label{eq:bggaugefields}
A_+^{bg} = (\Omega - W)^{-1} \Lambda_+, \qquad A_-^{bg} = (\mathbf{1} - W \Omega)^{-1} \Lambda_-  ,
\end{equation}
and, after passing to Euclidean signature, the tree-level contribution of the asymmetrical $\lambda$-model Lagrangian \eqref{eq:ActionLambdaCosetGaugeFields} on the background \eqref{eq:BGSSSM},\eqref{eq:bggaugefields} evaluates simply to,
\begin{equation}
L^0 (\lambda)  = \frac{k}{2\pi} \langle \Lambda_+ , (W \Omega + \mathbf{1}) (W\Omega - \mathbf{1})^{-1} \Lambda_- \rangle .
\end{equation}
To compute the one-loop contribution one  introduces a fluctuation around the background and integrates it out in the path integral by a saddle point approximation. Doing so, one needs to calculate the functional determinant of the operator that describes the equations of motion of the fluctuation.  Rather than carrying this out directly on the $\lambda$-model it is useful to observe that their equations of motion can be identified with those of the SSSM \eqref{eq:SSSMaction} where the computation is easier and described in detail in \cite{Appadu:2015nfa}.\\
\indent  To see this, let us consider the SSSM \eqref{eq:SSSMaction} and  define for now $\widehat{L}_\pm = \widehat{g}^{-1} \partial_\pm \widehat{g} - B_\pm$.  The equations of motion of the gauge field $B_\pm$ take the form of a constraint equation,
\begin{equation}
\widehat{L}^{(0)}_\pm = 0 .
\end{equation}
Subjected to this constraint,  the equations of motion and the Maurer-Cartan identity of the group-valued field $\widehat{g} \in G$ become, projected onto $\mathfrak{g}^{(0)}$ and $\mathfrak{g}^{(1)}$,
\begin{equation}\label{eq:EOMSSSM}
\begin{aligned}
&\partial_\pm \widehat{L}_\mp^{(1)} + [ B_\pm ,  \widehat{L}_\mp^{(1)} ] =0, \\
&\partial_+ B_- - \partial_- B_+ + [B_+, B_-] + [ \widehat{L}^{(1)}_+ ,  \widehat{L}^{(1)}_-] = 0 .
\end{aligned}
\end{equation}
One can, moreover, fix the gauge by a covariant gauge choice, 
\begin{equation} \label{eq:GCSSSM}
\partial_+ B_- + \partial_-  B_- = 0 .
\end{equation}
The equations of motion \eqref{eq:EOMSSSM}  can  be recast in terms of a flat Lax connection $\mathcal{L}(z)$, 
\begin{equation}
\mathcal{L}_\pm (z) = B_\pm + z^{\pm 1}  \widehat{L}^{(1)}_\pm ,
\end{equation}
satisfying $\left[ \partial_+ + \mathcal{L}_+(z), \partial_- + \mathcal{L}_-(z) \right] = 0$ for all $z\in \mathbb{C}$ and  ensuring the classical integrability of the SSSM. The SSSM Lax connection then indeed
 takes  an identical form to the Lax \eqref{eq:cosetlambdalax} of the  $\lambda$-deformed theory if we identify,
\begin{equation} \label{eq:identificationSSSMLambda}
\begin{aligned}
B_\pm  = - A^{(0)}_\pm, \qquad  \widehat{L}^{(1)}_\pm = - \lambda^{-1/2} A^{(1)}_\pm ,
\end{aligned}
\end{equation}
where the fields $A_\pm $  satisfy the constraints \eqref{eq:constraints}.\\
\indent For the one-loop contribution  we can now proceed with the SSSM as in section 2.2 of \cite{Appadu:2015nfa}  and subject the result to the identification \eqref{eq:identificationSSSMLambda}. Let us denote the   background fields  for  the gauge field $B_\pm$ and   the current $\widehat{L}^{(1)}_\pm$ by  $B^{bg}_\pm$ and $\Theta_\pm$ respectively, so that, 
\begin{equation}\label{eq:bgfields}
\begin{aligned}
B_\pm^{bg} &=0, \\
\Theta_+ &= - \lambda^{-1/2} (\Omega - W )^{-1} \Lambda_+ , \qquad \Theta_- &= - \lambda^{-1/2} ( \mathbf{1} -  W \Omega )^{-1} \Lambda_- ,
\end{aligned}
\end{equation}
where we assumed that $W$ respects the $\mathbb{Z}_2$-grading of $\mathfrak{g} = \mathfrak{g}^{(0)} \oplus \mathfrak{g}^{(1)}$ (as will be the case for the vector or axial deformed cases of section \ref{s:SL2U1})\footnote{When $W$ does not respect the $\mathbb{Z}_2$-grading one will generate non-zero background fields for the gauge fields $B_\pm$ and the calculation of \citep{Appadu:2015nfa} is not directly applicable anymore. In this case it seems that one needs to choose a different but appropriate background field for the group elements $g \in G$ than the one chosen in \eqref{eq:BGSSSM}. We will not consider this technical issue here further.}. Varying the equations of motion \eqref{eq:EOMSSSM} and the  covariant gauge fixing \eqref{eq:GCSSSM}  the operator that governs the fluctuations can be found, after Wick rotating to momentum space, to be,
\begin{equation}
\mathcal{D} = \begin{pmatrix}
p_- & 0 & 0 & -\Theta^{\text{adj}}_+ \\ 0 & p_+ & - \Theta^{\text{adj}}_- & 0 \\ -\Theta^{\text{adj}}_- & \Theta^{\text{adj}}_+ & -p_- & p_+ \\ 0 & 0 & p_- & p_+
\end{pmatrix},
\end{equation}
 acting on the fluctuations in the order $(\delta \hat{L}_+^{(1)},\delta \hat{L}_-^{(1)}, \delta B_+ , \delta B_-)$. Here we have $(\Theta^{\text{adj}}_\pm)_B{}^C =  \Theta_\pm^A (T_A^{\text{adj}})_B{}^{C} = i \Theta_\pm^A F_{AB}{}^{C} $.
The  one-loop contribution  to the Lagrangian, 
\begin{equation}
\begin{aligned}
L^1(\lambda) &= \frac{1}{2} \int^\mu \frac{\mathrm{d}^2 p}{(2\pi)^2} \Tr \log \mathcal{D},
\end{aligned}
\end{equation}
will have a logarithmic divergence given by \cite{Appadu:2015nfa}, 
\begin{equation}
\begin{aligned}
L^1(\lambda) = - \frac{c_2 (G)}{2\pi} \langle \Theta_+ , \Theta_- \rangle \log\mu + \cdots \\
\end{aligned}
\end{equation}
where $c_2(G) \equiv x_{\text{adj}}$ is the index of the adjoint representation. Substituting \eqref{eq:bgfields} and using the property \eqref{eq:Wconditions} that $W$ preserves the Lie algebra metric we find,
\begin{equation}
L^1(\lambda) = \frac{c_2 (G)}{2\pi} \frac{1}{\lambda} \langle \Lambda_+ , (W \Omega -  \mathbf{1})^{-1} W (W\Omega -  \mathbf{1})^{-1} \Lambda_- \rangle \log\mu + \cdots . 
\end{equation}
The one-loop beta function of the $\lambda$-parameter then follows from demanding that the one-loop effective Lagrangian $L(\lambda) =L^0(\lambda)+ L^1(\lambda)$ is independent of the scale $\mu$,
\begin{equation}
\mu \partial_\mu \left[ k \langle \Lambda_+ , \left( \frac{W\Omega + 1}{W \Omega - 1} \right) \Lambda_- \rangle  + \frac{c_2 (G)}{\lambda} \langle \Lambda_+ , (W\Omega - 1)^{-1} W (W\Omega - 1)^{-1} \Lambda_- \rangle \log\mu \right] = 0,
\end{equation}
 This yields (recall that $\Omega (\mathfrak{g}^{(1)}) = \lambda^{-1}$) to first order in $1/k$,
\begin{equation}
\mu \partial_\mu \lambda = - \frac{c_2 (G)}{2k} \lambda + \mathcal{O}\left(\frac{1}{k^2}\right) .
\end{equation}
We find   agreement with \citep{Appadu:2015nfa} and with \cite{Itsios:2014lca} for the case $G = SU(2)$, $H=U(1)$. We conclude that including an automorphism $W$ of the  Lie algebra $\mathfrak{g} = \mathfrak{g}^{(0)} \oplus \mathfrak{g}^{(1)}$ which respects the $\mathbb{Z}_2$-grading does not affect the one-loop beta function of the asymmetrical $\lambda$-model. As with the conventional symmetric $\lambda$-model, the deformation for compact groups is marginally relevant driving the model away from the CFT point   and marginally irrelevant for non-compact groups (as then one should send $k\rightarrow -k$, see appendix \ref{a:appendix}).


\subsection{Integrable boundary conditions}\label{s:IntBranes}
In this section we derive the (open string) boundary conditions that preserve integrability for the asymmetrical coset $\lambda$-model from the boundary monodromy method of \cite{Cherednik:1985vs,Sklyanin:1988yz,Dekel:2011ja,Driezen:2018glg}  to interpret them later as integrable D-brane configurations in the deformed background.

We define the generalised transport matrix,
\begin{equation}
\begin{aligned}
T^\mathcal{W}(b,a; z ) = \overleftarrow{P \exp} \left(- \int^b_a \mathrm{d}\sigma\; \mathcal{W}\left[ \mathcal{L}_\sigma (\tau,\sigma ; z)\right] \right)\, ,
\end{aligned}
\end{equation}
with an explicit dependence on the worldsheet coordinates $(\tau, \sigma$) included and where  $\mathcal{W}$ is a constant metric-preserving Lie algebra automorphism ($\mathcal{W}$ is not to be confused with the automorphism $W$ used in the asymmetric gauging). Generally speaking, under periodic boundary conditions (when $\partial\Sigma =0$) and with a flat Lax connection, one finds classical integrability by generating a tower of conserved charges from the \textit{monodromy matrix} $T^{\mathcal{W}}(2\pi,0;z)$ as $\partial_\tau \Tr T^{\mathcal{W}}(2\pi,0;z) ^n = 0$ with $n\in \mathbb{Z}$, see e.g. \cite{Babelon:2003qtg}. This is not the case under open boundary conditions. Instead, we build the \textit{boundary monodromy matrix} $T_b (z)$ by gluing the usual ($\mathcal{W}=\mathbf{1}$) transport matrix $T(\pi,0;z)$ (from the $\sigma = 0$ to the $\sigma = \pi$ end) to the generalised transport matrix $T_R^{\mathcal{W}}(2\pi,\pi;z)$ in the reflected region:
\begin{equation}
 \begin{aligned}
 T_b (z) = T_R^{\mathcal{W}}(2\pi,\pi;z)  T(\pi,0;z),
 \end{aligned}
 \end{equation} 
 where $T_R^{\mathcal{W}}(2\pi,\pi;z)$ is constructed from the Lax \eqref{eq:cosetlambdalax} under the reflection $\sigma \rightarrow 2\pi - \sigma$ so that,
 \begin{equation}
 T_R^{\mathcal{W}}(2\pi,\pi;z) = T^{\mathcal{W}}(0,\pi; z^{-1}).
 \end{equation}
 One finds an infinite set of conserved charges given by $\Tr T_b(z) ^n = 0$ with $n\in \mathbb{Z}$  when $\partial_\tau T_b (z) = \left[ T_b(z), N(z) \right]$ for some $N(z)$. This is satisfied sufficiently  when $N(z) = \mathcal{L}_\tau (0;z) $ and when we impose the  boundary conditions  \cite{Dekel:2011ja,Driezen:2018glg}:
 \begin{equation} \label{eq:intbc}
\left.  \mathcal{L}_\tau ( z) \right\vert_{\partial\Sigma}  = \left.   \mathcal{W}\left[ \mathcal{L}_\tau (  z^{-1} ) \right] \right\vert_{\partial\Sigma} ,
 \end{equation}
 on both the open string ends. Explicitly, for the Lax connection \eqref{eq:cosetlambdalax} of the $\lambda$-coset model, we find by expanding order by order in the arbitrary parameter $z$ the conditions,
 \begin{subequations}\label{eq:gluingcondmain}
 \abovedisplayshortskip=0pt
\abovedisplayskip=0pt
 \begin{alignat}{2}
& \mathcal{O}(z): \qquad  & \left. A^{(1)}_+ \right\vert_{\partial\Sigma} = \left. \mathcal{W} [ A_-^{(1)} ] \right\vert_{\partial\Sigma} ,  \label{eq:gluingcond}\\
 &\mathcal{O}(z^0): \qquad  & \left. A^{(0)}_\tau \right\vert_{\partial\Sigma} = \left. \mathcal{W} [ A_\tau^{(0)} ] \right\vert_{\partial\Sigma} ,\label{eq:concond1} \\
 &\mathcal{O}(z^{-1}): \qquad  & \left. A^{(1)}_- \right\vert_{\partial\Sigma} = \left. \mathcal{W} [ A_+^{(1)} ] \right\vert_{\partial\Sigma}  \label{eq:gluingcond2}   .
 \end{alignat}
 \end{subequations}
 Note from the above  that the automorphism $\mathcal{W} $ should respect the $\mathbb{Z}_2$ grading.
 Moreover, from \eqref{eq:concond1} one deduces that $\mathcal{W}(\mathfrak{g}^{(0)}) = \mathbf{1}$  unless $A_\tau^{(0)}|_{\partial \Sigma} = 0$ and using \eqref{eq:gluingcond2} in \eqref{eq:gluingcond} that $\mathcal{W}^{2}(\mathfrak{g}^{(1)})= \mathbf{1}$.   Taking these restrictions on $\mathcal{W}$ into account we continue with \eqref{eq:gluingcond} as describing the \textit{integrable} boundary conditions. In components, and using the constraint equations \eqref{eq:constraints}, it translates to conditions on the local coordinates $X^{\mu}$ as,
\begin{equation} \label{eq:intbcexpl}
\left. \left[(D_g W - \Omega)^{-1} \right]^\alpha{}_B R^B{}_\mu \partial_+ X^{\mu} \right\vert_{\partial\Sigma} = \left. - \mathcal{W}^{\alpha}{}_\beta \left[(D_{g^{-1}} - W \Omega  )^{-1} \right]^\beta{}_C L^C{}_\mu \partial_- X^\mu \right\vert_{\partial\Sigma}.
\end{equation} 
Given a $G/H$ model one can now continue by studying the eigensystem and derive the corresponding D-brane configurations in the target space background. This will be illustrated in section \ref{s:CigarBranes} for $G = SL(2,R)$ and $H=U(1)$.\\
\indent In \cite{Driezen:2018glg} we described also the possibility to glue $T(\pi, 0 ;z)$ to a gauge transformed reflected transport matrix $T_R^{{\cal W} g}(2\pi, \pi ;z)$. Here we have the residual gauge symmetry \eqref{eq:residualtransf} under which the Lax \eqref{eq:cosetlambdalax} transforms as ${\cal L}(z) \rightarrow h^{-1}{\cal L} h + h^{-1}\mathrm{d} h$ with $h\in H$. The integrable boundary conditions  then read,
 \begin{equation}\label{eq:intbcgauge}
 \left. {\cal L}_\tau (  z ) \right\vert_{\partial\Sigma} = \left.  {\cal W} \left[h^{-1} {\cal L}_\tau ( z^{-1}) h  + h^{-1} \partial_\tau h \right]  \right\vert_{\partial\Sigma} ,
 \end{equation}
which  allows a gluing of the gauge fields that is field-dependent. We  will see in the explicit example of section \ref{s:SL2U1} that this possibility will prove to be of significant importance to exhibit distinct D-brane configurations. 



\section{Deforming the Euclidean black hole and Sine-Liouville} \label{s:SL2U1}

We now illustrate the general story above with a simple example.  The simplest example one could consider is the $SU(2)/U(1)$ case, however, there are no non-trivial outer automorphisms here and all that is achieved is simply a coordinate redefinition as seen from \eqref{eq:AsymmetricGaugingFields}.  One could go on to look at compact theories based on  e.g. $SU(3)$ which does have such a symmetry however we choose here instead to pursue directly the $SL(2,R)/U(1)$ theories given their interest towards black hole physics.

For $G = SL(2,R)$ we take our generators $T_A$, $A=\{1,2,3\}$ to be,
\begin{equation}
T_1 =\frac{1}{\sqrt{2}} \begin{pmatrix}
1 & 0 \\ 0 & -1
\end{pmatrix}, \quad T_2 = \frac{1}{\sqrt{2}} \begin{pmatrix}
0 & 1 \\ 1 & 0
\end{pmatrix}, \quad T_3 = \frac{1}{\sqrt{2}} \begin{pmatrix}
0 & 1 \\ -1 & 0
\end{pmatrix},
\end{equation}
such that $\Tr (T_A  T_B ) = \text{diag}(+1,+1,-1)$ and adopt the following  parameterisation of a group element $g \in SL(2,R)$ connected to the identity,
\begin{equation}
\begin{aligned}
g = e^{\frac{\tau-\theta}{\sqrt{2}}\,T_3}\,e^{\sqrt{2}\,\rho\, T_1}\,e^{\frac{\tau+\theta}{\sqrt{2}}\,T_3} = \cosh\rho  \begin{pmatrix}
\cos\tau & \sin\tau \\ -\sin\tau & \cos\tau
\end{pmatrix} + \sinh\rho \begin{pmatrix}
\cos\theta & \sin\theta  \\ \sin\theta  & -\cos\theta 
\end{pmatrix} \ , 
\end{aligned}
\end{equation}
with $\rho\in [0, + \infty ) $,  $\theta ,\tau \in [0,2\pi] $. We take the subgroup $H=U(1)$ to be generated by $T_3$.

\subsection{The parafermionic $SL(2,R)/U(1)$ WZW theory}

Let us first consider gauging the $U(1)_k$ subgroup in the WZW model on (a single cover of) $SL(2,R)_k$. As a coset CFT this model can be understood as being generated by a set of non-compact \textit{parafermionic} currents introduced in \cite{Lykken:1988ut} which are semi-local chiral fields with fractional spin (see also \cite{Bakas:1991fs} and for the compact analogues \cite{Fateev:1985mm}). In terms of these \cite{Bakas:1991fs} showed the symmetry algebra to be  the non-linear infinite W-algebra $\hat{W}_\infty(k)$. Although obscured as a non-rational CFT it is expected that, as in  the compact $SU(2)/U(1)$  theory \cite{Fateev:1985mm,Maldacena:2001ky}, the level $k$ parafermion theory and its ${\mathbb Z}_k$ orbifold are equivalent for $k$ integral \cite{Dijkgraaf:1991ba,Israel:2003ry}.

For large $k$ we can view these theories as sigma models  for strings propagating in a two-dimensional target space equipped with a non-constant dilaton originating from the action \eqref{eq:ASGWZW}.
%
%
%
%
 If we perform an axial gauging $g \rightarrow h g h $ with $h \in H$ the $\tau$-coordinate is gauge and we obtain, up to finite $1/k$ corrections, the \textit{cigar} geometry,
\begin{equation} \label{eq:axialWZWSL2U1}
\mathrm{d}s^2_A = k \left( \mathrm{d}\rho^2 + \tanh^2 \rho \, \mathrm{d}\theta ^2 \right) , \quad e^{-2\Phi_A} = e^{-2\Phi_0}\cosh^2 \rho ,
\end{equation}
and zero B-field. The geometry is semi-infinite and terminates at $\rho = 0$ where the dilaton field is of maximum but finite value. The Ricci scalar computed from this metric is $R = \frac{4}{k \cosh^2\rho}$ so that  $\rho = 0 $ is only a coordinate singularity.   \\
\indent  If instead we perform the vector gauging $g \rightarrow  h^{-1} g h$ the coordinate $\theta $ is gauge and we find at large $k$ the \textit{trumpet} geometry,
\begin{equation}\label{eq:vectorWZWSL2U1}
\mathrm{d}s^2_V = k \left( \mathrm{d}\rho^2 + \coth^2 \rho \, \mathrm{d}\tau^2 \right) , \quad e^{-2\Phi_A} = e^{-2\Phi_0}\sinh^2 \rho ,
\end{equation}
and zero B-field. The Ricci scalar is now $R = - \frac{4}{k \sinh^2\rho}$ and, therefore, $\rho = 0$ is a true curvature singularity where the dilaton field reaches $+\infty$. Notice that both solutions \eqref{eq:axialWZWSL2U1}  and \eqref{eq:vectorWZWSL2U1} are related by the  transformation,
\begin{equation}
\begin{aligned}
\rho &\rightarrow \rho + \frac{i \pi}{2}, \qquad \theta  \rightarrow \tau .
\end{aligned}
\end{equation}
which, because it involves a complexification, is obviously not a standard field redefinition. Below we will understand it as originating from an outer automorphism.  When performing an analytical continuation to Lorentzian signature the above solutions can be interpreted as a two-dimensional black hole for which the global Kruskal coordinates were written down in \cite{Witten:1991yr}. The cigar and trumpet solutions correspond to the region outside the horizon and inside the singularity respectively and are described by an equivalent coset CFT \cite{Dijkgraaf:1991ba} with a central charge,
\begin{equation} \label{eq:centralchargeSL2U1}
c = \frac{3k}{k-2} -1 \ .
\end{equation}
As we will see shortly, the cigar  is known to be T-dual to the $\mathbb{Z}_k$ orbifold  of the trumpet solution, and vice versa, where in the Euclidean picture the orbifolding can be understood as  changing the temperature of the black hole \cite{Dijkgraaf:1991ba,Giveon:1991sy,Kiritsis:1991zt,Rocek:1991ps}.

The axial gauged   $SL(2,R)/U(1)$ WZW \eqref{eq:axialWZWSL2U1}  has a  $U(1)_\theta $ isometry shrinking to zero size at $\rho = 0$ breaking the conservation of winding number. Nevertheless one can associate a classically  conserved current  $J_\pm^\theta $ to $U(1)_\theta $ given by,
\begin{equation}\label{eq:phiconserv}
J_\pm^\theta =  k \tanh^2\rho \partial_\pm\theta , \qquad  \partial_+ J_-^\theta  + \partial_- J_+^\theta  =0 .
\end{equation}
Using the  conservation equation together with the equations of motion for $\rho, \theta$, one can give semi-classical analogues of the non-compact parafermions  which furnish chiral algebra's,
\begin{equation}
\partial_- \Psi^{A}_{(\pm)} = \partial_+ \bar{\Psi}^{A}_{(\pm)}  = 0 ,
\end{equation}
in terms of phase space variables \cite{Bardacki:1990wj,MariosPetropoulos:2005rtq},
\begin{equation} \label{eq:AxialParafermions}
\begin{aligned}
\Psi^{A}_{(\pm)} &= \left( \partial_+ \rho \mp i \tanh\rho \partial_+\theta  \right) e^{\mp i (\theta  + \frac{\tilde{\theta}}{k})}, \qquad
  \bar{\Psi}^{A}_{(\pm)} = \left( \partial_- \rho \pm i \tanh\rho \partial_- \theta  \right) e^{\pm i (\theta  - \frac{\tilde{\theta}}{k})} ,
\end{aligned}
\end{equation}
where $\tilde{\theta}$ is a non-local expression in terms of $\rho$ and $\theta $ defined by,
\begin{equation} \label{eq:canTdSL2U1}
\partial_\pm \tilde{\theta} = \pm J_\pm^\theta  .
\end{equation}
This relation corresponds precisely to the canonical T-duality rule  found when performing a standard Buscher procedure  \cite{Buscher:1987sk,Buscher:1987qj,Alvarez:1994wj} on the $U(1)_\theta $ isometry. In the dual picture  $\tilde{\theta }$ becomes a local coordinate with a periodicity of $2\pi$ \cite{Rocek:1991ps}. The T-dual background is,
\begin{equation}\label{eq:OrbifoldSL2U1}
\mathrm{d}s^2_{O} = k \left( \mathrm{d}\rho^2 + \frac{1}{k^2} \coth^2 \rho \mathrm{d}\tilde{\theta } \right) , \quad e^{-2\Phi_{O}} = e^{-2\Phi_0} \sinh^2\rho,
\end{equation}
and thus corresponds to the $\mathbb{Z}_k$ orbifold of the vectorial gauged theory  \eqref{eq:vectorWZWSL2U1}. Acting with the T-duality action \eqref{eq:canTdSL2U1} the non-compact parafermions of the dual background become,
\begin{equation}\label{eq:TdRuleOnParafermions}
\begin{aligned}
\Psi^{A}_{(\pm)} &\rightarrow \Psi^{O}_{(\pm)} = \left( \partial_+ \rho \mp i \coth\rho \frac{\partial_+\tilde{\theta } }{k}\right) e^{\mp i (  \frac{\tilde{\theta }}{k} + \theta )}, \\
 \bar{\Psi}^{A}_{(\pm)} &\rightarrow  \bar{\Psi}^{O}_{(\pm)} = \left( \partial_- \rho \mp i \coth\rho \frac{\partial_-\tilde{\theta } }{k} \right) e^{\mp i (  \frac{\tilde{\theta }}{k} - \theta)} ,
 \end{aligned}
\end{equation}
in which now $\theta$ is a non-local expression in the fields $\rho$ and $\tilde{\theta}$ satisfying,
\begin{equation}
\partial_\pm \theta = \pm J^{\tilde{\theta }}_\pm, \qquad  J^{\tilde{\theta }}_\pm = \coth^2 \rho  \frac{\partial_\pm\tilde{\theta}}{k}  , 
\end{equation}
with $J^{\tilde{\theta}}_\pm$ the $U(1)_{\tilde{\theta }}$ classically conserved current of the background  \eqref{eq:OrbifoldSL2U1}. Together with the classical equations of motions, this ensures again the dual parafermions  to be holomorphically conserved, $\partial_- \Psi^{O}_{(\pm)}  = \partial_+  \bar{\Psi}^{O}_{(\pm)} = 0$.

\subsection{Asymmetrical $\lambda$-deformed $SL(2,R)/U(1)$}

Let us now consider the asymmetrically deformed $\lambda$-theories. The metric preserving automorphisms $W$  satisfying \eqref{eq:Wconditions} are  elements of $SO(2,1)$ (including elements disconnected from the identity). They can for instance act as,
\begin{equation}
W : \{ T_1 , T_2, T_3 \} \mapsto \{ T_1, \cosh\alpha T_2 + \sinh\alpha T_3 , \sinh\alpha T_2 + \cosh\alpha T_3 \} , 
\end{equation}
induced from the action on $g\in SL(2,R)$ by $g \mapsto w g w^{-1}$ with,
\begin{equation}\label{eq:groupauto}
w = \exp\left( \frac{\alpha}{\sqrt{2}} T_1 \right) .
\end{equation}
When the parameter $\alpha \in \mathbb{R}$ the asymmetric gauging involves an  inner automorphism which from \eqref{eq:AsymmetricGaugingFields} can clearly be absorbed by a trivial field redefinition.
When instead we take for instance $\alpha = i \pi$ we have $w \in SL(2,\mathbb{C})$ and hence the automorphism $W$ is outer. It is an element of $SO(2,1)$  corresponding to a reflection of the $T_2$ and $T_3$ directions (i.e.\ $W = \text{diag}(+1,-1,-1)$) and is thus disconnected from the identity.  The corresponding asymmetrical $\lambda$-theory then defines  a background that deforms the axial gauged $SL(2,R)/U(1)$ WZW (since $W(T_3) = -T_3$) or \textit{cigar} geometry of \eqref{eq:axialWZWSL2U1}. Under the residual gauge symmetry \eqref{eq:residualtransf} the $\tau$-coordinate is then indeed gauge so that we can adopt the  gauge fixing choice $\tau = 0$. Introducing the complex coordinates $\zeta = \sinh\rho e^{i \theta} $ and $\bar{\zeta} = \sinh\rho e^{-i \theta} $ the group element can then be written as,
 \begin{equation}\label{eq:gfix}
\begin{aligned}
g &=\left( \begin{array}{cc}
 \cosh \rho +\cos \theta  \sinh \rho  & \sin \theta  \sinh \rho  \\
 \sin \theta  \sinh \rho  & \cosh \rho -\cos \theta  \sinh \rho  \\
\end{array}  
\right),\\
 &=  \frac{1}{2}\left(
\begin{array}{cc}
 \zeta + \bar{\zeta} -2 \sqrt{\zeta  \bar{\zeta} +1} &
   - i (\zeta - \bar{\zeta} ) \\
 - i (\zeta - \bar{\zeta} ) & -\zeta - \bar{\zeta} -2
   \sqrt{\zeta   \bar{\zeta} +1} 
   \end{array}
\right) .
\end{aligned}
\end{equation} 
The gauge field equations of motion \eqref{eq:constraints} are, 
\begin{equation}\label{eq:gaugeeqm}
\begin{aligned}
& (1- \lambda) A_+^1 + i (1+\lambda)A_+^2 = -\frac{\sqrt{2} \lambda}{\sqrt{1+ \zeta \bar\zeta} } \partial_+ \zeta ,  \\ 
& (1- \lambda) A_-^1 + i (1+\lambda)A_-^2 =  \frac{\sqrt{2} \lambda}{\sqrt{1+ \zeta \bar\zeta}}  \partial_- \bar\zeta ,
\end{aligned}
\end{equation} 
with $A^3_\pm$ determined in terms of $A^1_\pm$ and $A^2_\pm$.  The deformed background can be computed from \eqref{eq:LambdaActionEffective} and \eqref{eq:LambdaDilaton} to be,
 \begin{equation} \label{eq:defcigar1}
\begin{aligned}
\mathrm{d}s^2_{A,\lambda} &= k \left( \frac{1-\lambda}{1+\lambda} \left(\mathrm{d}\rho^2 + \tanh^2\rho \mathrm{d}\theta ^2 \right) + \frac{4 \lambda}{1-\lambda^2} \left(\cos\theta  \mathrm{d}\rho - \sin\theta  \tanh\rho \mathrm{d}\theta  \right)^2  \right),\\
&= \frac{k}{1-\lambda^2} \frac{ \left( \lambda \left(\mathrm{d}\zeta^2 + \mathrm{d}\bar{\zeta}^2 \right) + (1+\lambda^2)\mathrm{d}\zeta \mathrm{d}\bar{\zeta}  \right)}{1+|\zeta |^2}, \\
e^{-2\Phi} &= e^{-2\Phi_0}\cosh^2 \rho = e^{-2\Phi_0} \left(1+ |\zeta |^2  \right),
\end{aligned}
\end{equation}
and zero B-field. Notice that the deformation has broken the $U(1)_\theta$ isometry  to a $\mathbb{Z}_2$. As before, $\rho =0$ is only a coordinate singularity  where the dilaton is constant.\\
\indent Note that for $\lambda = 0$ we have that the metric is of the form $\mathrm{d}s^2_A = k \partial \bar \partial V(\zeta \bar\zeta) d\zeta d \bar\zeta $ with $V(x) = -\text{Li}_2(-x) =  \int_0^x ds s^{-1} \log(1+s)$ and the geometry  is indeed K\"ahler \cite{Rocek:1991vk} allowing ${\cal N} = (2,2)$ worldsheet supersymmetry.  Let us see if we can find a similar form in the deformation, i.e.\ as $\mathrm{d} s^2_{A,\lambda} =  k \partial \bar\partial V^\lambda(\zeta, \bar{\zeta}) d\zeta d \bar\zeta$, with an eye on future applications to extended worldsheet supersymmetry.   First, let us bring the metric into canonical form by defining $\zeta  =   Z - \lambda \bar{Z}$  such that,
\begin{equation}
\mathrm{d} s^2_{A,\lambda} =  k \frac{(1-\lambda^2) d Z d \bar Z }{  1 - \lambda (Z^2 + \bar{Z}^2) + (1+\lambda^2) Z \bar Z },
\end{equation}
Although performing directly a double integral of  the function $ (1+\lambda^2) ( 1 - \lambda (Z^2 + \bar{Z}^2) + (1+\lambda^2) Z \bar Z )^{-1}$  appears to be inaccessible  one can however do an expansion in $\lambda$ and integrate each term in this evolution.  To first order we find,
\begin{equation}
V^\lambda(Z, \bar{Z}) = -\text{Li}_2(- Z \bar{Z}) + \lambda \left(\frac{1}{Z^2} + \frac{1}{\bar{Z}^2} \right) \log (1 + Z \bar{Z}) - \lambda \left( \frac{Z}{\bar{Z}} + \frac{\bar{Z}}{Z} \right) + {\cal O}(\lambda^2 ).
\end{equation}
Whilst a series expansion can doubtless be found, the resummation of such a result is not evident.  However, this first-order perturbed potential can be the starting point for the development of the notion of integrability in an ${\cal N} = (2,2)$ superspace setting, a totally uncharted topic. We hope to come back to this in a future publication.  \\
%
%
%
%
%
%
%
%
%
\indent For the remains of the paper we will see it to be more useful to reformulate the deformation in terms of the axial parafermions \eqref{eq:AxialParafermions}. The Lagrangian $L_A$ of the sigma model corresponding to the deformed  geometry \eqref{eq:defcigar1} is a perturbation of the CFT point $L_{A,\mbox{\tiny{WZW}}}$  by a  bilinear in the axial parafermions (as in \cite{Sfetsos:2013wia}) given to all orders by,
\begin{equation}\label{eq:AxialSmallLambda}
L_A =   \ k \left( \frac{1+\lambda^2}{1-\lambda^2} L_{A,\mbox{\tiny{WZW}}} + \frac{\lambda}{1-\lambda^2} ( \Psi^{A}_{(+)} \bar{\Psi}^{A}_{(-)} + \Psi^{A}_{(-)} \bar{\Psi}^{A}_{(+)} )  \right) .
 \end{equation} 
Notice that the non-local phases  $\tilde{\theta}$ of the parafermions drop out of this bilinear combination. Furthermore, this perturbation is clearly a non-compact analogue of the one considered in \cite{Fateev:1991bv}.

\indent When instead we take $\alpha = 0$ in \eqref{eq:groupauto} and thus $W$ the identity (that is trivially inner) one obtains the background known from \cite{Sfetsos:2014cea}, or from an analytical continuation of the $SU(2)/U(1)$ case of \cite{Sfetsos:2013wia},
\begin{equation} \label{eq:deftrumpet1}
 \begin{aligned}
 \mathrm{d}s^2_{V,\lambda} &= k \left(\frac{1-\lambda}{1+\lambda} \left(\mathrm{d}\rho^2  + \coth^2\rho \mathrm{d}\tau^2   \right) +  \frac{4\lambda}{1-\lambda^2} \left( \cos\tau \mathrm{d}\rho - \sin\tau\coth\rho \mathrm{d}\tau \right)^2 \right), \\
 e^{-2\Phi} &= e^{-2\Phi_0} \sinh^2 \rho,
 \end{aligned}
 \end{equation}
 and zero B-field, deforming the  vectorial gauged \textit{trumpet} geometry of \eqref{eq:vectorWZWSL2U1}. Here $\rho = 0$ is again representing the curvature singularity\footnote{After analytical continuation, reference \cite{Sfetsos:2014cea} derived the global Kruskal coordinates of the vectorially deformed theory  to interpret the background as a deformed two-dimensional black hole capturing therefore also the region outside the horizon. However, a systematic analysis to obtain this region from an axial gauged deformation  was lacking there.}. After taking the $\mathbb{Z}_k$ orbifold, where the coordinate $\tau$ is replaced by the  $2\pi/k$ periodic coordinate $\tilde{\theta}/k$, the first order correction to the corresponding Lagrangian $L_O$ becomes a bilinear in terms of the  orbifold parafermions  $\Psi_{\pm}^{O}$ of \eqref{eq:TdRuleOnParafermions} as \cite{Sfetsos:2013wia},
 \begin{equation} \label{eq:VectorSmallLambda}
   L_O =  \ k  \left(\frac{1+\lambda^2}{1-\lambda^2} L_{O,\mbox{\tiny{WZW}}} + \frac{\lambda}{1-\lambda^2}   ( \Psi^{O}_{(+)} \bar{\Psi}^{O}_{(+)} + \Psi^{O}_{(-)} \bar{\Psi}^{O}_{(-)} )  \right) ,
 \end{equation}
 in which again the non-local phases drop out. One might at first sight think this indicates the axial-vector duality of the CFT point ($\lambda = 0$)  \cite{Dijkgraaf:1991ba,Giveon:1991sy,Kiritsis:1991zt,Rocek:1991ps} to persist in the deformation. However, one needs to be more careful here: when performing the T-duality transformation \eqref{eq:TdRuleOnParafermions} on \eqref{eq:AxialSmallLambda} the $\Psi_{(\pm)}^{O}$ enter in a  combination where the non-local $\theta$ does not drop out and so the   deformation term \eqref{eq:VectorSmallLambda} is not recovered.    Indeed this can be expected as the  deformation  destroys the isometries of the background.
 

%
 
\subsection{Integrable branes in the $\lambda$-cigar} \label{s:CigarBranes}

Let us now consider  integrable boundary conditions defined  in  the $\lambda$-cigar geometry. Even in the undeformed case, this is a challenging question because of the well known difficulties with non-rational CFT. However, the expectation is (and based on a semi-classical analysis of the DBI axtion) that the cigar geometry allows D0-, D1- and D2-brane configurations \cite{Fotopoulos:2003vc,Ribault:2003ss,Fotopoulos:2004ut,Israel:2004jt,Ribault:2005pq}. Except for the D0, these branes can be understood as descending from the ungauged $SL(2,R)$ WZW model \cite{Bachas:2000fr}. Geometrically, the D0  is located at the tip of the cigar, the D1 covers a so-called \textit{hairpin} and the D2 is either space-filling or extends from the circle at some value $\rho_\star >0$  to infinity. The D1-branes are understood to be non-compact analogues of the A-branes of \cite{Maldacena:2001ky} in the $SU(2)/U(1)$ WZW while the D0 and D2 are analogues of the B-branes. The latter are an interesting type as they  provide a way to derive symmetry breaking branes  in the parent theory which are non-obvious to obtain from first principles, see for instance \cite{Quella:2002ct} and references therein. Here we will find the above D-brane configurations by employing the classical  integrability technique outlined in section \ref{s:IntBranes}.

\begin{figure}[H]
\centering
\includegraphics[scale=0.35]{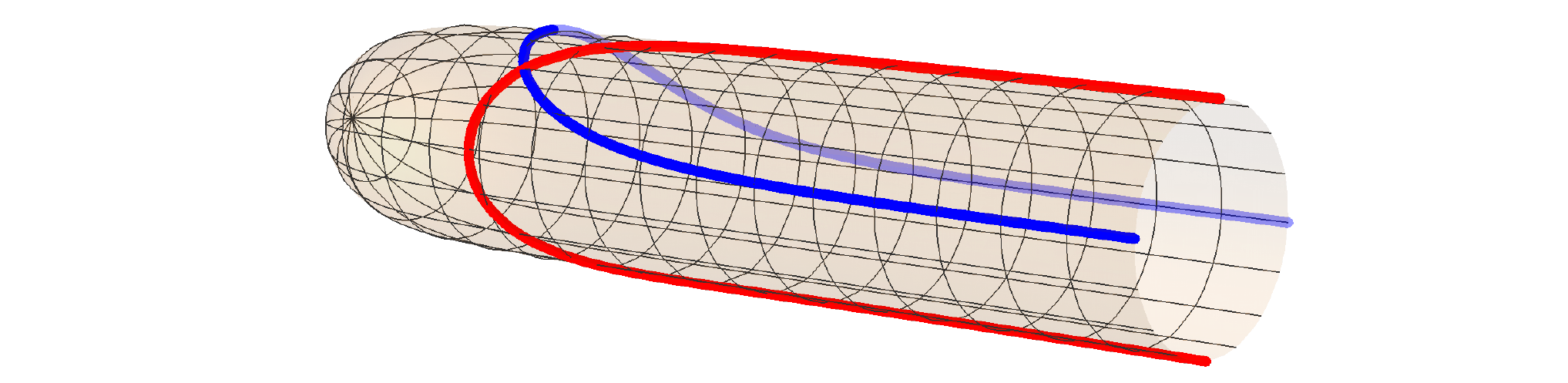}
\caption[protect]{The D1-brane configurations in the undeformed cigar manifold embedded in $\mathbb{R}^3$. Heuristically, one can think of the deformation as to convert the $U(1)_\theta$ circle into an ellipse. However, visualising this exactly is surprisingly challenging\protect\footnotemark.}\label{fig:CigarBranes}
\end{figure}

We start with analysing  the simplest case given in equations (\ref{eq:intbc},~\ref{eq:intbcexpl})   for the cigar, i.e.\ taking $W=  \text{diag}(1,-1,-1)$, and for $\mathcal{W}=\mathbf{1}_3$ (which is trivially satisfying the restrictions given below \eqref{eq:gluingcondmain}). After a straightforward computation this leads to the integrable boundary conditions,
\begin{equation}
\begin{aligned}
\cos\theta \partial_\tau \rho - \sin\theta\tanh\rho \partial_\tau\theta &= 0,\\
\sin\theta \partial_\sigma \rho + \cos\theta\tanh\rho \partial_\sigma \theta &=0 ,
\end{aligned}
\end{equation}
which describe static D1-branes. These boundary conditions notably do not depend on the deformation parameter and indeed match precisely those of the CFT point \cite{Fotopoulos:2003vc,Ribault:2003ss,Fotopoulos:2004ut}. 
In terms of the complex coordinates $\zeta = \sinh\rho e^{i \theta} $, $\bar{\zeta} = \sinh\rho e^{-i \theta} $ they simplify to,
\begin{equation}\label{eq:D1}
\begin{aligned}
\partial_\tau \left( \zeta + \bar{\zeta} \right) &= 0,  \quad
\partial_\sigma \left( \zeta - \bar{\zeta} \right) &=0 .
\end{aligned}
\end{equation}
The Dirichlet condition gives the embedding equation in the two-dimensional $(\rho, \theta)$ space such that the D1-branes cover so-called \textit{hairpins} on the cigar as visualised in figure
\ref{fig:CigarBranes} in the undeformed case. In the  limit $\rho \rightarrow \infty$ the branes reach the asymptotic circle at two opposite positions, $\theta = \pi/2, 3\pi/2$. Another possibility in the $\lambda$-cigar is taking the gluing automorphism $\mathcal{W} = \text{diag}(-1,-1,1)$. In this case the integrable boundary conditions \eqref{eq:intbcexpl} are an exchange of the Dirichlet and Neumann direction,
\begin{equation}\label{eq:D1rot}
\begin{aligned}
\partial_\tau \left( \zeta - \bar{\zeta} \right) &= 0, \quad 
\partial_\sigma \left( \zeta + \bar{\zeta} \right) &=0 ,
\end{aligned}
\end{equation}
corresponding to a  rotation along the circle of the static D1-branes over an angle $\pi/2$. In contrast to the undeformed case, the extra restrictions on the automorphism $\mathcal{W}$ prevents the  branes to  be  rotated smoothly into each other while preserving the integrability properties, essentially since the deformation destroys such isometry of the background.
\indent Let us  consider the D1-branes found above also from the semi-classical perspective. If we let $y$ be the spatial coordinate of the D1-brane\footnotetext{Whilst it is easy to find an explicit isometric embedding in $\mathbb{R}^3$ for the undeformed cigar geometry, finding the same for the deformed cigar proved to be an engrossing, deceptively challenging, and ultimately frustrating activity, at least for the present authors.  Solutions to this problem would be welcomed.}\footnote{As is commonplace in the topic we assume that there is an auxiliary time direction and assume some static gauge.} and introduce $u = |\zeta | = \sinh(\rho)$ then the  DBI action reads, 
\begin{equation}
\begin{aligned}
S_{DBI} = T_1 \int \mathrm{d}y \, e^{-\Phi} \sqrt{\det \hat{G}},
   \end{aligned}   
\end{equation}
where,
\begin{equation}
\begin{aligned}
e^{-2\Phi} \det\hat{G} & \propto u'(y)^2 \left(1+ \lambda ^2+2  \lambda  \cos (2 \theta (y))\right)  -4 \lambda  u(y) u'(y) \theta '(y) \sin (2 \theta (y)) \\
   & \qquad +u(y)^2 \theta '(y)^2 \left(1+ \lambda ^2-2 \lambda
    \cos (2 \theta (y))\right) . 
       \end{aligned}   
\end{equation}
Although the action evidently depends on the deformation parameter, this drops out in the classical Euler-Lagrange equations, which have a solution,  
\begin{equation}
u(y)=\upsilon  \csc ( \theta_0+\theta (y)),
\end{equation}
with $\upsilon, \theta_0$ integration constants.  Hence, the D1-branes are semi-infinite with $u \in (\upsilon, \infty)$.  Plugging this solution back into the DBI action yields, 
\begin{equation}
S_{DBI} \propto \lim_{u\to\infty}  \sqrt{u^2-\upsilon ^2} \sqrt{1+\lambda ^2+2 \lambda  \cos (2  \theta_0 ) } \ . 
\end{equation}
Whilst this is clearly diverging, for any UV cut-off the action is minimised by $\theta_0 = \frac{\pi}{2} ,\frac{3\pi}{2}$. Asymptotically as $\rho \to \infty$ these special configurations match precisely to the integrable D-branes described in \eqref{eq:D1}.

As is the case in the undeformed cigar we anticipate\footnote{Inspired by \cite{Driezen:2018glg} where a generic geometrical approach was taken for group manifolds, we anticipate the brane configurations of the CFT to persist in the deformed theory.} here also $D0$-branes localised at the tip.  The corresponding worldsheet boundary conditions read, 
\begin{equation}
\partial_\tau \theta = \partial_\tau \rho = 0 \ , \quad \rho = 0 \ .
\end{equation} 
To ascertain if these constitute integrable boundary conditions we shall reverse the logic compared to the $D1$ case described above; we shall start with these boundary conditions on the field and from this infer a boundary condition on the Lax connection.  A first step is to use the gauge field equations eq.~\eqref{eq:gaugeeqm} of motion evaluated with the gauge fixing choice eq.~\eqref{eq:gfix}.  Then the D0 boundary condition reads simply, 
\begin{equation}
A^1_+ = A^1_-  \ , \quad  A^2_+ = - A^2_- , \quad  A^3_+ =   A^3_- = 0 \ , 
\end{equation} 
where the latter equality follows on $\rho = 0$.
In terms of the Lax connection \eqref{eq:cosetlambdalax},
\begin{equation}
{\cal L}_\tau(z) = \frac{1}{\sqrt{2 \lambda} z}  \left(
\begin{array}{cc}
 -(1+ z^2 )A_+^1 &
 (1 -z^2) A_+^2 \\
(1 -z^2) A_+^2&
 (1+z^2 )A_+^1\\
\end{array}
\right)
\end{equation}
 we find that this satisfies the condition $ \left. {\cal L}_\tau(z) \right\vert = \left. {\cal W} [{\cal L}_\tau(z^{-1}) ] \right\vert$ of   \eqref{eq:intbc} when ${\cal W}= \textrm{diag}(1,-1,-1)$.  In this case ${\cal W}$ satisfies all necessary requirements when $\rho = 0$ (since then $A_\tau^3 = 0$): it is a constant metric-preserving automorphism of $\mathfrak{sl}(2,R)$ and ${\cal W}^2(\mathfrak{g}^{(1)}) = \mathbf{1}$.  

In \cite{Fotopoulos:2003vc,Ribault:2003ss} it was shown that there is also a D2-brane configuration  supported by a worldvolume gauge field ${\cal A}$ with field strength $F_{\rho \theta}\equiv f=   \partial_\rho {\cal A}_\theta$  (in which the gauge ${\cal A}_\rho=0$ is adopted).  In the deformed scenario we might again anticipate finding such a configuration.  Indeed from the DBI action, 
\begin{equation}
S_{DBI} \propto \int d\rho d\sigma e^{-\Phi} \sqrt{ \det(G + F)} , 
\end{equation} 
we find that the $\lambda$-dependence drops from the equation of motion for the gauge field which is solved with, 
\begin{equation}
f^2 = \frac{ \beta^2 \tanh^2 \rho } {-\beta^2 + \cosh^2\rho} \ .
\end{equation}
Here we see that when the constant $\beta >1$, the field strength $f$ is critical outside the region $\cosh\rho \geq \beta$ so that the D2-brane extends from the asymptotic circle to a minimum value in $\rho$ given by $\cosh\rho_\star = \beta$. When $\beta<1$, however, the D2 is space-filling.\\
\indent The question now comes if this corresponds to an integrable boundary condition.  Recall that a volume-filling brane should consist of generalised Neumann type boundary conditions that incorporate the gauge field $F$: 
\begin{equation}
G_{ab} \partial_\sigma X^a = F_{ab} \partial_\tau X^b \ . 
\end{equation}
In terms of the coordinates $X= (\rho ,\theta)$ these are quite inelegant and have explicit dependance on $\lambda$.  However, we may recast this result in terms of the  gauge fields $A^{(1)}_\pm$ using the on-shell equations of motion \eqref{eq:gaugeeqm}. We find that upon doing so the $\lambda$-dependence  is again removed and yields, 
\begin{equation} \label{eq:D2bcgaugefields}
(1 +f^2 \coth^2 \rho) \{ A^1_- , A^2_-\} =  (1 -f^2 \coth^2 \rho) \{ -A^1_+ ,A^2_+\} - 2 f \coth \rho \{ A^2_+, A^1_+ \} .
\end{equation}
This tells us the gluing between the gauge fields should be field-dependent and therefore hints towards a boundary condition of the form \eqref{eq:intbcgauge} where one includes a gauge transformation in the boundary monodromy matrix. Indeed, after a tedious but straightforward computation we find that gauge transforming the Lax \eqref{eq:cosetlambdalax},
\begin{equation}
{\cal L} (z) \rightarrow h^{-1} {\cal L} (z) h + h^{-1} \mathrm{d}h,
\end{equation}
by,
\begin{equation}
h =  \exp \left(  v(\rho,\beta)    T_3 \right) \in H , \qquad v(\rho,\beta) = \sqrt{2} \arcsin \left( \coth^2 \rho f^2 +1  \right)^{-1/2},
\end{equation}
 the integrable boundary condition \eqref{eq:intbcgauge} agrees with the D2 boundary conditions \eqref{eq:D2bcgaugefields} when $\mathcal{W} = \text{diag}(1,-1,-1)$.
 
 Concluding, we see here  integrable D-branes corresponding to D0-, D1- and D2-configurations which are all obtained differently from a boundary condition on the Lax connection. We see also that not all of the D1-branes of the undeformed theory  preserve integrability: instead of having the continuous $U(1)_\theta$ isometry, only two configurations at specific angles  survive the integrable deformation.

\subsection{Connection to Sine-Liouville theory}

 We are now in a position to discuss  the deformation to the dual Sine-Liouville (SL) background, which in the undeformed case has the action (see for instance \cite{Kazakov:2000pm,Fateev:2017mug}), 
\begin{equation}\label{eq:actionSineLiouville}
S_{SL,k} (x,\phi) = \frac{1}{\pi} \int_\Sigma d\tau d\sigma \;   \partial_+ \phi \partial_- \phi + \partial_+ x \partial_- x + Q R^{(2)} \phi + \mu e^{b\phi} \cos (R \tilde{x}) ,
\end{equation}
with $R^{(2)}$  the worldsheet Ricci scalar. The target space has the topology of cylinder with $\phi \in (-\infty,+\infty)$  the radial coordinate and $x$  a $2\pi$ periodic coordinate with radius $R$ and a dual $\tilde{x}$. The parameters $Q$, $b$ and $R$ are related as $Q = -1/b$ and $R^2-b^2 = 2$ ensuring Sine-Liouville is an exact CFT with central charge,
\begin{equation} \label{eq:centralchargeSL}
c = 2 + 6 Q^2 ,
\end{equation}
and a potential $V(\phi,\tilde{x}) = \mu e^{b\phi} \cos (R \tilde{x})$ with scaling dimension $1$.
The central charge of the Euclidean cigar \eqref{eq:centralchargeSL2U1} matches with that of SL when $Q^2 = \frac{1}{k-2}$, hence (taking the positive root of $Q$) we have $b =- \sqrt{k-2}$ and  $R = \sqrt{k}$.\\
\indent A dictionary between the (undeformed) Euclidean cigar black hole and Sine-Liouville theory  can be made in the asymptotic flat space limit $\rho \rightarrow \infty$ where the cigar approaches the toplogy of a cylinder and its dilaton falls off linearly, $\Phi_A - \Phi_0 \rightarrow - \rho$. On the SL side, this limit corresponds to the region $\phi \rightarrow \infty$ in which the potential $V(\phi,\tilde{x})$ as well as the string coupling constant go to zero given the dilaton $\Phi_{SL} =  Q \phi$.   The identification is therefore at large $k$ given by,
\begin{equation} \label{eq:CigarVsSL}
\rho \sim - Q \phi, \qquad \theta \sim \frac{x}{\sqrt{k}}  , \qquad  \tilde{\chi} \sim \sqrt{k} \tilde{x} \ . 
\end{equation}
At finite $\rho$ and $\phi$, the duality between both theories can be demonstrated as an exact match between the symmetry algebra's, vertex operators and $n$-point functions \cite{Fateev:SL,Kazakov:2000pm,Hikida:2008pe} (see also \cite{Fateev:2017mug}) where they look both topologically and dynamically  very different. Indeed, it can be understood that the dynamics is governed by the geometry in the cigar picture and by  the potential $V(\phi,\tilde{x})$ in the  SL picture. Additionally, the tip of the cigar is the end of space corresponding to the horizon of the Euclidean black hole and hence cutting off the strong string coupling region, while on the SL side this region is protected by the potential $V(\phi,\tilde{x})$. On the worldsheet the duality can be viewed as a strong-weak coupling duality. However, the sigma model point of view taken here forces us  in the small coupling (large $k$) regime on the cigar side.\\
\indent For us the power of the duality lies in the observation  that the semi-classical cigar parafermions \eqref{eq:AxialParafermions} in the flat space limit under the identification \eqref{eq:CigarVsSL}, 
\begin{equation}\label{eq:SLParafermions}
\begin{aligned}
\Psi_{(\pm)}^{SL} &= \left( - \frac{ \partial_+ \phi}{\sqrt{k-2}} \mp i \frac{\partial_+ x}{\sqrt{k}} \right) e^{\mp \frac{2i x_L}{\sqrt{k}}}, \qquad
\bar{\Psi}_{(\pm)}^{SL} = \left( - \frac{ \partial_- \phi}{\sqrt{k-2}} \pm i \frac{\partial_- x}{\sqrt{k}} \right) e^{\pm \frac{2i x_R}{\sqrt{k}}}, 
\end{aligned}
\end{equation}
commute\footnote{After analytical continuation to Euclidean worldsheet signature one should check that $\oint_w \mathrm{d}z \Psi_{(\pm)}^{SL}(z) V(\phi(w),\tilde{x}(w)) =
\oint_{\bar{w}} \mathrm{d}\bar{z} \bar{\Psi}_{(\pm)}^{SL}(\bar{z}) V(\phi(\bar{w}),\tilde{x}(\bar{w})) 
$. Note that a translation to  \cite{Fateev:2017mug} should be done in the large $k$ limit and by  the substitution $\phi \rightarrow \varphi /2$, $x \rightarrow \phi / 2$, $b\rightarrow 2 b$, $R\rightarrow 2 a$. Doing so one indeed finds $\Psi^{SL}_{(\pm)} \propto \Psi^{\text{Fateev}}_{(\mp)}$ up to an irrelevant overall factor.} with the SL  potential $V(\phi,\tilde{x})$ \cite{Fateev:2017mug}. Here $x (\sigma^+ , \sigma^- ) = x_L (\sigma^+ ) + x_R (\sigma^-)$ and  $\tilde{x} (\sigma^+ , \sigma^- ) = x_L (\sigma^+ ) - x_R (\sigma^-)$. Therefore, one can  rely on the expression \eqref{eq:SLParafermions} for all values of $\phi$.  Since the parafermion fields induce the deformation  \eqref{eq:AxialSmallLambda}  we can now easily extract the perturbation on the SL theory side. To first order in $\lambda$ the deforming term in the large $k$ regime becomes,
 \begin{equation}
 \begin{aligned}
 \delta L_{SL} = \lambda &\left( 2 \cos\left(\frac{2 x}{R} \right) \partial_+ \phi \partial_- \phi - 2 \cos\left(\frac{2 x}{R} \right)    \partial_+ x\partial_- x  \right. \\
& \;\;\left.    + 2  \sin\left(\frac{2 x}{R} \right)  (\partial_+ x \partial_- \phi + \partial_- x \partial_+ \phi ) \right) + {\cal O}(\lambda^2) \ .
 \end{aligned}
 \end{equation}
A similar structure is expected for finite $\lambda$, as \eqref{eq:AxialSmallLambda} is exact in $\lambda$, so that one deforms the flat space SL theory to a curved background. We anticipate this is the starting point of an integrable deformation of the SL theory. Moreover, it appears to be in a different class to the integrable deformations studied in \cite{Fateev:2017mug}. We will leave this as an open problem to be fully understood.


\section{Conclusion} \label{s:conclusions}

The Sfetsos procedure \cite{Sfetsos:2013wia} to construct the $\lambda$-deformation of a $G/H$ coset realised as a gauged WZW model actually requires the $G/G$ model as a starting point.  To date, even when $H$ is abelian,  attention has been restricted to the case in which in the $G/G$ model the $G$ symmetry, and  consequently that of $H$, acts vectorially.   Here we explore the asymmetric gauging of $G$ in which the left and right actions differ by the application of an algebra automorphism.  When this is an outer automorphism what results can not be trivially removed via field redefinitions.   In this way,  we are able to produce  new $\lambda$-type deformations leading to topologically distinct target spaces in a robust and fundamental manner. Using the similarities between this asymmetric $\lambda$-model and its vectorial cousin we demonstrate classical integrability  and show the one-loop beta functions to stay marginally relevant for compact groups and irrelevant for non-compact groups. To end our general discussion of this model, we present a simple technique to  construct  integrable boundary conditions in which we, moreover, exploit the residual asymmetric gauge symmetry. \\
\indent As an example we consider the $SL(2,R)/U(1)$ model where unlike the compact $SU(2)$ there is such a non-trivial outer automorphism.  We show that employing our procedure we are able to find an integrable deformation of the theory in which the gauged symmetry acts axially.  Geometrically, and at large $k$, we have an integrable deformation of the cigar geometry corresponding to the Euclideanised Witten black hole. The cigar geometry itself receives $\frac{1}{k}$ corrections and it  would be doubtless valuable to find a description of the $\lambda$-deformation that takes these corrections into account.   Continuing at large $k$, we analyse also the boundary conditions preserving integrability in the deformed cigar. We see this  can be done  straightforwardly and observe the D-branes proposed at the (non-rational) CFT point to be integrable in the deformation.

As well as demonstrating the concept for this broader class of deformations we believe this example could hold some further interest in its own right.  Let us entertain some speculation about how the deformation translates to both the Sine-Liouville (SL) dual and in turn to the matrix model description of this picture.   An initial step is made here by identifying for small deformation parameters in the cigar a bilinear of the non-compact parafermions as the operators that drive the deformation.  Demanding agreement between the SL at large values of the radial coordinate suggests strongly the same parafermionic bilinear deformation should be considered in the SL model.  However the $\lambda$-model goes much further since it provides a resummation to all orders in $\lambda$ of this deformation; what this looks like in the SL theory is far from clear.  One possible root to shed light on this could be to combine the Sfetsos procedure with the path integral derivation of FZZ. When successful, one can continue and probe, using the deformed SL theory and integrability, the region behind the horizon. \\
\indent It is also interesting to ask what the deformation does at the level of the S-matrix.  For the case of similar deformations of {\em compact} parafermionic theories it has long been known that the S-matrix has a kink structure and in the $k \to \infty$ limit matches that of the $O(3)$ sigma-model \cite{Fateev:1991bv}.  A similar expectation holds for general $\lambda$-deformations, the underlying S-matrix has a $q$ root-of-unity quantum group symmetry associated to a face model \cite{Hollowood:2015dpa,Appadu:2017fff}.  Here it is less clear due to the non-compactness of the theory but one might well anticipate a similar $q$-deformation to hold.   Further one might ask what this structure might relate to in the postulated dual matrix model description of the cigar  \cite{Kazakov:2000pm}.  \\
\indent A final enticing direction is to employ similar techniques in the context of geometries relevant to black hole microstates.  For instance a static configuration of NS5-branes on a circle admits a description as a gauged WZW model \cite{Sfetsos:1998xd,Israel:2004ir}, and more general solutions (supertubes and spectral flows of supertubes)  can also be realised as gauged WZW models \cite{Martinec:2017ztd,Martinec:2018nco}.  It seems quite possible that the techniques developed here may be applicable to such situations.  We leave that for future work.

\section*{Acknowledgments}

\noindent
  We thank  Ben Hoare,  Tim Hollowood, Carlos Nunez and Kostas Sfetsos  for useful discussions that aided this project and to Panagiotis Betzios, Gaston Giribet, Olga Papadoulaki and David Turton for useful communications on the manuscript.    DCT is supported by a Royal Society University Research Fellowship {\em Generalised Dualities in String Theory and Holography} URF 150185 and in part by STFC grant ST/P00055X/1. SD is supported by the ``FWO-Vlaanderen'' through an aspirant fellowship. This work is additionally supported in part by the ``FWO-Vlaanderen'' through the project G006119N and by the Vrije Universiteit Brussel through the Strategic Research Program ``High-Energy Physics''.  

%
%
%
%
%
%
%
%

\appendix

\section{Conventions and sigma models (WZW, PCM and SSSM)} \label{a:appendix}

In this appendix, we briefly introduce some basic ingredients and conventions for the gauging procedure of section \ref{s:construction}. 

 For the general formulae of this paper we adopt conventions for compact and semi-simple groups $G$, although they should be changed conveniently when working out the non-compact $SL(2,R)/U(1)$ example in section \ref{s:SL2U1}. We denote the generators of the Lie algebra $\mathfrak{g}$ of  $G$ by $T_A$ and pick a basis in which they are Hermitean, i.e. $\left[ T_A , T_B \right] = i F_{AB}{}^C T_C$  with real structure constants $F_{AB}{}^C$ and $A = \{ 1, \cdots , \dim G \}$. They are normalised in such a way that the ad-invariant Cartan-Killing metric $\langle \cdot , \cdot \rangle : \mathfrak{g} \times \mathfrak{g} \rightarrow \mathbb{R}$, taken to be $\langle T_A , T_B \rangle = \frac{1}{x_R} \Tr \left(T_A T_B \right)$ with  $x_R$ the index of the representation $R$,  has unit entries. The left-(right-)invariant Maurer-Cartan one-forms are expanded in the Lie algebra as $g^{-1}\mathrm{d} g = -i L^A T_A $ ($\mathrm{d}g g^{-1} = -i R^A T_A$) and in  explicit local  coordinates $X^{\mu}$, $\mu\in\{1, \cdots , \dim G\}$ as $g^{-1} \mathrm{d} g = -i L^A{}_\mu(X) T_A  \mathrm{d}X^\mu$ ($ \mathrm{d}g g^{-1} = -i R^A{}_\mu(X) T_A \mathrm{d}X^\mu $). The adjoint action is denoted by $D_g T_A = g T_A g^{-1} = (D_g)^B{}_A T_B$, hence $(D_g)_{AB} = \langle  T_A  ,g  T_B g^{-1} \rangle$ and $R^A = (D_g)^A{}_B L^B$.  \\
\indent Finally, considering the $G/H$ coset, we denote the  generators of the subgroup $H \subset G$  with Lie algebra $\mathfrak{h}$ by $T_a$, $a =\{1, \cdots, \text{dim}H \}$  and the remaining generators  by $T_\alpha$, $\alpha = \{ \text{dim}H + 1, \cdots, \text{dim}G\}$. We assume the Lie algebra $\mathfrak{g}$ to have a symmetric space decomposition $\mathfrak{g} = \mathfrak{g}^{(0)} \oplus \mathfrak{g}^{(1)}$, with $\mathfrak{g}^{(0)} \equiv \mathfrak{h}$, defined by a  $\mathbb{Z}_2$ grading $[\mathfrak{g}^{(i)}, \mathfrak{g}^{(j)}] \subset \mathfrak{g}^{(i+j \text{ mod }2)}$.

We consider the WZW model on a Lie group manifold $G$ at level $k$  \cite{Witten:1983ar} with the action,
\begin{equation}
S_{\mbox{\tiny{WZW}},k}(g) = - \frac{k}{2\pi} \int_{\Sigma} d\sigma d\tau \langle  g^{-1}\partial_+ g , g^{-1} \partial_- g \rangle - \frac{k}{24\pi} \int_{M_3} \langle \bar{g}^{-1} \mathrm{d}\bar{g} , \left[  \bar{g}^{-1} \mathrm{d}\bar{g} ,  \bar{g}^{-1} \mathrm{d}\bar{g} \right]  \rangle ,
\end{equation}
with $g : \Sigma \rightarrow G$ a Lie group element and  $\bar{g}$ an extension of $g$ into $M_3 \subset G$ such that $\partial M_3 = g(\Sigma)$. To cancel ambiguities from the choice of $M_3$ in the path integral the level $k$ should be integer quantised for compact groups while for non-compact cases it can be free   \cite{Witten:1983ar,Figueroa-OFarrill:2000lcd}. The two-dimensional   manifold $\Sigma$  can be thought of as a worldsheet on which we have fixed  the metric as  $\text{diag}(+1,-1)$, the Levi-Civita as $\epsilon_{\tau\sigma} = 1$ and we have units in which  $\alpha' = 1$ .  We analytically continue to Euclidean coordinates by taking $ \sigma_+ = \tau + \sigma \to -i z $ and $\sigma_- = \tau - \sigma \to -i \bar{z} $ and will use the term \textit{holomorphic} abusively to mean either $f(\sigma^+)$ or $f(z)$.  The WZW model on group manifolds is known to have an exact CFT formulation originating from the $G_L(\sigma^+) \times G_R(\sigma^-)$ symmetry generated by the  holomorphically conserved currents $J_+ (\sigma^+) = - k \partial_+ g g^{-1}$ and $J_- (\sigma^-) = k g^{-1} \partial_- g$ whose components satisfy two commuting Kac-Moody algebra's. \\
\indent We consider moreover the PCM model on a Lie group manifold $G$  with a coupling constant $\kappa^2$,
\begin{equation}\label{ActionPCM}
S_{\mbox{\tiny{PCM}}, \kappa^2} (\widehat{g}) = - \frac{\kappa^2}{\pi} \int d\sigma d\tau \langle \widehat{g}^{-1}\partial_+ \widehat{g} ,  \widehat{g}^{-1}\partial_- \widehat{g} \rangle , \quad \widehat{g} \in G ,
\end{equation}
which has a global $G_L \times G_R$ symmetry. 
From the PCM model  the SSSM model on the $G/H$ coset manifold can be obtained by gauging  an $H_R \subset G$ subgroup acting as,
\begin{equation}
\widehat{g} \rightarrow \widehat{g} h.
\end{equation}
 The gauge-invariant action is then,
\begin{equation}
S_{\mbox{\tiny{SSSM}},\kappa^2}(\widehat{g}, B_\pm)  = - \frac{\kappa^2 }{\pi} \int d\sigma d\tau \langle (\widehat{g}^{-1}\partial_+ \widehat{g} - B_+),( \widehat{g}^{-1}\partial_- \widehat{g} - B_-) \rangle ,
\end{equation}
with  $B_\pm$ the gauge fields taking values in the Lie algebra  $\mathfrak{g}^{(0)} \equiv \mathfrak{h}$ of $H$ and transforming under the gauge transformation as $ B_\pm \rightarrow h^{-1} \left( B_\pm  + \partial_\pm \right) h$. This model is easily shown to be classically integrable when $\mathfrak{g} = \mathfrak{g}^{(0)} \oplus \mathfrak{g}^{(1)}$ has a symmetric space decomposition \cite{Eichenherr:1979ci,Hollowood:2014rla}.

Note that when working with non-compact groups, where one picks an anti-Hermitean basis to have real structure constants, one should analytically continue in the above models $k \rightarrow - k$ and $\kappa^2 \rightarrow - \kappa^2$ in order to keep the right sign on the kinetic term.

\bibliographystyle{JHEP}
\bibliography{SibBib}

\end{document}